\documentclass[11pt,a4paper]{article}

\usepackage[utf8]{inputenc}
\usepackage[T1]{fontenc}
\usepackage{geometry}
\usepackage{amsmath,amssymb}
\usepackage{graphicx}
\usepackage{hyperref}
\usepackage{url}
\usepackage{booktabs}
\usepackage{xcolor}
\usepackage{authblk}  %
\usepackage{multirow} %
\usepackage{threeparttable}  %
\usepackage[numbers,square]{natbib}  %
\usepackage{subcaption} %

\newcommand{\received}[1]{}  %
\newcommand{\revised}[1]{}   %
\newcommand{\accepted}[1]{}  %
\newcommand{\journal}[1]{}   %
\newcommand{\volume}[1]{}    %
\newcommand{\copyyear}[1]{}  %
\newcommand{\startpage}[1]{} %

\usepackage[nolist]{acronym} 
\usepackage{todonotes}
\setuptodonotes{inline}
\usepackage{textgreek}
\usepackage{natbib}  %
\begin{document}

\title{Exploring Definitions of Quality and Diversity in Sonic Measurement Spaces}

\author[1,2]{Björn Þór Jónsson\thanks{bthj@uio.no}}
\author[2]{Çağrı Erdem}
\author[3]{Stefano Fasciani}
\author[1,2]{Kyrre Glette}

\affil[1]{RITMO, University of Oslo, Oslo, Norway}
\affil[2]{Department of Informatics, University of Oslo, Oslo, Norway}
\affil[3]{Department of Musicology, University of Oslo, Oslo, Norway}

\renewcommand\Authfont{\fontsize{10}{12}\selectfont}
\renewcommand\Affilfont{\fontsize{9}{10.8}\itshape}

\maketitle

\begin{abstract}
Digital sound synthesis presents the opportunity to explore vast parameter spaces containing millions of configurations. Quality diversity (QD) evolutionary algorithms offer a promising approach to harness this potential, yet their success hinges on appropriate sonic feature representations. Existing QD methods predominantly employ handcrafted descriptors or supervised classifiers, potentially introducing unintended exploration biases and constraining discovery to familiar sonic regions. This work investigates unsupervised dimensionality reduction methods for automatically defining and dynamically reconfiguring sonic behaviour spaces during QD search. We apply Principal Component Analysis (PCA) and autoencoders to project high-dimensional audio features onto structured grids for MAP-Elites, implementing dynamic reconfiguration through model retraining at regular intervals. Comparison across two experimental scenarios shows that automatic approaches achieve significantly greater diversity than handcrafted behaviour spaces while avoiding expert-imposed biases. Dynamic behaviour-space reconfiguration maintains evolutionary pressure and prevents stagnation, with PCA proving most effective among the dimensionality reduction techniques. These results contribute to automated sonic discovery systems capable of exploring vast parameter spaces without manual intervention or supervised training constraints.
\end{abstract}

\footnote{\textbf{Abbreviations:} QD, Quality Diversity; DR, Dimensionality Reduction; CPPN, Compositional Pattern Producing Network.}

\section{Introduction}\label{secIntro}

Modern digital audio synthesis systems generate sounds by navigating high-dimensional parameter spaces that span millions of possible configurations. However, systematically exploring these vast sonic territories remains challenging given the inherent difficulties in characterising and navigating such high-dimensional spaces. Manual approaches require domain expertise and risk overlooking novel sonic relationships, whilst supervised machine learning methods remain constrained by their training data, potentially limiting exploration to familiar sonic territories.

Evolutionary algorithms offer a promising approach to automated sonic discovery, but their effectiveness 
critically depends 
on how they characterise and navigate the underlying sonic behaviour space. 
\ac{QD} algorithms like \ac{MAP-Elites} \cite{mouret_illuminating_2015} have shown success in various domains, yet their application to sonic discovery has typically relied on manually crafted behaviour descriptors or supervised classifiers that may impose unintended biases on the exploration process.

This work addresses the need for automated, unbiased methods to define and dynamically reconfigure sonic behaviour spaces for evolutionary search. We investigate two core research questions: \textbf{Do automatic behaviour space definition approaches outperform manual behaviour space definition for sonic quality diversity search in terms of 
the 
diversity 
of discovered sounds?} This encompasses different \ac{DR} methods 
and dynamic reconfiguration strategies compared to expert-selected features. \textbf{How do different quality evaluation approaches affect the effectiveness of sonic discovery?} This examines reference-based quality assessment (single and Multiple Reference) versus reference-free evaluation methods.

We investigate the application of unsupervised \ac{DR} methods to automatically define and dynamically reconfigure sonic behaviour spaces during evolutionary search, adapting approaches from robotics research to the sonic domain. Specifically, we apply 
the 
\ac{DR} techniques 
\ac{PCA} and autoencoders to project high-dimensional audio features onto structured two-dimensional grids for \ac{MAP-Elites} operation. To prevent evolutionary stagnation, we implement dynamic behaviour space reconfiguration through periodic retraining of these models using sounds discovered during the search process.

Our experimental results show that unsupervised \ac{DR} models can autonomously discover meaningful sonic diversity without expert bias, achieving significantly higher diversity than manually crafted behaviour spaces. The contributions to quality diversity algorithms applied to audio synthesis are: (1) demonstrating that unsupervised behaviour space definition, including dynamic reconfiguration, achieves substantially higher sonic diversity than expert-defined spaces whilst maintaining sound quality; and (2) comparing reference-based quality evaluation approaches (single and multiple reference sounds) with reference-free methods, revealing how different quality assessment strategies affect the diversity and perceptual characteristics of discovered sounds.

The following sections review related work in evolutionary sound synthesis and quality diversity algorithms, detail our methodology for unsupervised behaviour space definition, present experimental results, and discuss implications for automated sonic discovery systems.

\section{Related Work}\label{secRelatedWork}

\subsection{Evolutionary Computation for Sound Synthesis}
Efforts to expand our sonic vocabulary through evolutionary algorithms are wide and varied. One branch of these efforts has involved searching the parameter spaces of established sound synthesisers \cite{horner_machine_1993, tatar_automatic_2016}, while another has focused on evolving the structure of sound generators themselves \cite{manzolli_evolutionary_2001, macret_automatic_2014}. More recent explorations have ventured into the latent spaces of deep generative models \cite{guo_lvns-rave_2024}.

Early pioneering work established fundamental approaches to evolutionary sound synthesis, including FM parameter matching \cite{horner_machine_1993} and the evolution of sound synthesis algorithms \cite{manzolli_evolutionary_2001}. Comprehensive surveys by Romero et al. \cite{romero_evolutionary_2008} documented the breadth of evolutionary approaches to sound synthesis, while interactive evolution has enabled exploration of complex parameter spaces through human aesthetic guidance \cite{dahlstedt_creating_2001, jonsson_interactively_2015}.

Human input has been incorporated into evolutionary processes through interactive evolutionary computation \cite{takagi_interactive_2001}, though our initial explorations into applying this approach to the discovery of sounds \cite{jonsson_interactively_2015} revealed how challenging that domain can be for human evaluators. The limited capacity of human evaluation \cite{hutchison_open_2005} is particularly noticeable in sound evaluation, where mental fatigue can arise from repeatedly assessing diverse sounds. Those limitations make it especially 
appealing to 
seek for approaches that 
enable automatic evolutionary search for sounds.

\subsection{Quality Diversity Algorithms and Innovation Engines}

Drawing inspiration from the accumulation of all interesting discoveries in nature and human culture,
\ac{QD} search algorithms \cite{pugh_quality_2016, chatzilygeroudis_quality-diversity_2021} drive evolution towards novelty whilst also taking into account measures of quality or fitness. These methods shift the focus from finding a single optimum to producing collections of high-quality solutions that together illuminate the space of possibilities \cite{stanley_why_2015}. \ac{MAP-Elites}, our chosen algorithm, organises search into a structured, grid-based container where each cell stores the best solution exhibiting a given behaviour, thereby encouraging broad coverage across behavioural dimensions rather than collapse to a single peak.

\ac{QD} algorithms have demonstrated success across diverse creative applications beyond robotics, including procedural content generation and game level design \cite{gravina_procedural_2019,khalifa_talakat_2018,liapis_transforming_2013}, demonstrating how these approaches can illuminate design spaces in creative contexts and reveal possibilities that traditional optimisation methods might overlook.

Building upon \ac{QD} algorithms, the \textit{Innovation Engine} concept \cite{nguyen_innovation_2015, nguyen_understanding_2016} combines diversity-promoting evolutionary search with learned representations that both define behavioural spaces and evaluate solution quality. In initial implementations, pre-trained deep neural networks provided both the structure of the behaviour space (through class labels) and the fitness signal (through classification confidence), enabling automated creative discovery at scales impossible with interactive evolution. This configuration has succeeded in producing interesting visual \cite{nguyen_understanding_2016, lehman_creative_2016} and sonic \cite{jonsson_towards_2024} artefacts. However, these implementations remain constrained by what can be characterised within their training data: pre-trained classifiers guide discovery towards artefacts recognisable within predefined taxonomies of categories, potentially limiting exploration to familiar territories whilst overlooking genuinely novel concepts that don't fit established classifications.

The conceptual foundations of Innovation Engines can be traced through a progression of key developments. Novelty search \cite{lehman_abandoning_2011} demonstrated that rewarding behavioural novelty could outperform objective-based approaches in deceptive environments. Interactive evolutionary systems like Picbreeder \cite{secretan_picbreeder_2011} showed how human aesthetic guidance could lead to serendipitous discoveries, though with scalability limitations due to user fatigue. \ac{QD} algorithms, particularly \ac{MAP-Elites}, provided structured methods for maintaining diverse collections whilst illuminating fitness landscapes. The Innovation Engine concept united these insights by automating aesthetic judgement through learned representations whilst preserving open-ended exploration. Further research has investigated whether \ac{QD} algorithms provide better stepping stones for exploration compared to objective-based search \cite{gaier_are_2019}, whilst recent extensions explore automated creativity through deep learning integration \cite{aki_llm-poet_2024}.

The ultimate vision for Innovation Engines would operate without requiring labelled data or domain-specific knowledge \cite{nguyen_understanding_2016}. Our work explores an alternative approach to this vision by decoupling behavioural characterisation from quality evaluation. Rather than deriving both from a single learned model, we employ unsupervised dimensionality reduction techniques (\ac{PCA} and autoencoders) to define behaviour spaces autonomously, whilst quality evaluation operates through independent mechanisms including reference-based similarity measures and reference-free audio quality assessment. This separation enables exploration beyond the categorical constraints of supervised classifiers, potentially revealing sonic territories that pre-trained models might not recognise or value.

\subsection{Behaviour Space Definition in Quality Diversity}\label{sec:behaviour_space_definition}
Search based on the family of \ac{QD} algorithms relies on characterisations referred to as \acfp{BD}. A collection of such behaviours describes characteristics of the artefacts the algorithm searches for, be that robotic behaviours, images, or sounds. Traditional approaches have relied on manually defined behaviour descriptors or supervised classifiers, where manual definition requires domain expertise and may miss important relationships, whilst supervised approaches can inherit biases from their training data.

\ac{QD} algorithms like \ac{MAP-Elites} highlighted the importance of appropriate behaviour characterisation through systematic studies \cite{pugh_confronting_2015}, which emphasised how the alignment between quality measures and behaviour descriptors impacts algorithmic performance. Recent work in robotics has begun exploring autonomous characterisation through unsupervised descriptors. AURORA \cite{cully_autonomous_2019, grillotti_unsupervised_2022} employs dimensionality reduction models (encoders) trained on sensory data from discovered solutions to automatically define behavioural characterisations. The algorithm operates with unstructured containers where diversity of characteristics is illuminated through distance comparisons in high-dimensional latent spaces. 
Complementary approaches have explored unsupervised learning of reachable outcome spaces \cite{paolo_unsupervised_2020} and robust \ac{QD} operation in noisy domains typical of robotic applications \cite{flageat_fast_2020}. In contrast with those previous studies, we investigate the suitability of illuminating a diversity of sound objects within dynamically structured two-dimensional containers.

\subsection{Dynamic Behaviour Space Reconfiguration}
Static behaviour spaces may become saturated during evolution, potentially limiting continued discovery. Expansive variants of \ac{MAP-Elites} \cite{vassiliades_comparison_2017} address this limitation through dynamic reconfiguration, typically by expanding container sizes or introducing new behavioural dimensions.

AURORA (introduced in Section~\ref{sec:behaviour_space_definition}) advances this approach through periodic retraining of its dimensionality reduction models \cite{cully_autonomous_2019, grillotti_unsupervised_2022}. The algorithm alternates between a \ac{QD} phase, where solutions fill an unstructured container, and an encoder update phase, where the DR model is incrementally retrained on sensory data from discovered solutions. This periodic redefinition allows the behavioural space to adapt to the evolving distribution of discoveries whilst maintaining continuity through incremental parameter updates. To manage container size in these unbounded spaces, adaptive distance thresholds determine when solutions are sufficiently novel to be added to the container.

Alternative approaches have explored transformational creativity through incremental learning of new behavioural representations \cite{liapis_transforming_2013}, where autoencoders progressively refine the criteria for novelty evaluation. Our work adapts these periodic redefinition principles to the sonic domain with structured grid-based containers, as detailed in Section~\ref{secMethod}.

\subsection{Quality Diversity in Sound Matching vs Sound Discovery}
\ac{QD} algorithms have been applied to different sonic contexts. While this work focuses on discovering diverse sonic artefacts through open-ended exploration, other research has applied \ac{QD} algorithms to the problem of sound matching. Masuda \& Saito \cite{masuda_quality-diversity_2023} introduced novelty objectives to genetic algorithm-based synthesiser parameter estimation, seeking multiple diverse ways to recreate a target sound rather than a single optimal match. Their approach employs Novelty Search with Global Competition (NS-GC) and Local Competition (NS-LC) variants, using manually selected behavioural descriptors (spectral centroid and spectral flatness) to characterise sonic diversity.

These two applications represent 
different contexts for \ac{QD} algorithms. Sound matching approaches employ reference sounds as specific targets to guide search toward particular sonic characteristics, whilst sound discovery approaches employ reference sounds as quality thresholds for exploring diverse sonic territories. The former defines desired output characteristics through reference targets; the latter establishes minimum coherence criteria whilst permitting broad exploration across the behaviour space. Sound matching approaches often benefit from manual selection of acoustically meaningful behavioural characteristics, whilst discovery approaches may explore automated behaviour space definition. Fixed synthesiser architectures (such as subtractive and FM synthesis) provide interpretable parameter spaces suited to matching tasks, whilst open-ended generative systems allow for structural evolution at the cost of reduced interpretability.

The distinction between \textit{sound matching} and \textit{sound discovery} reflects broader applications in creative AI systems. Matching problems typically benefit from expert knowledge of relevant acoustic features and well-defined target characteristics. Discovery problems often explore automated characterisation approaches to avoid constraining exploration to familiar territories, as different applications may warrant different balance points between expert guidance and automated exploration.

\subsection{Compositional Pattern Producing Networks for Sound}
The sound generation approach employed in our experiments builds upon the application of evolutionary algorithms to sound synthesis through \acp{CPPN} \cite{stanley_compositional_2007}. Those networks have been developed to represent or abstract unfolding development in evolutionary processes, which build a phenotype over time, and this can be compared with the process of timbral development, where musical expression depends on changes and nuances over time.

CPPNs have demonstrated remarkable versatility across multiple domains, from the evolution of visual patterns in collaborative systems like Picbreeder \cite{secretan_picbreeder_2011} to recent applications in automated environment generation \cite{aki_llm-poet_2024}. In the sonic domain, CPPNs have been applied to interactive sound synthesis \cite{jonsson_interactively_2015}, demonstrating their capacity to evolve complex timbral structures through neuroevolution.

Our distinct approach to sound synthesis couples \acp{CPPN} with \ac{DSP} graphs \cite{jonsson_system_2024}, to both of which we apply \ac{NEAT}. This design poses challenges as it does not encode any inductive bias towards recognisable sounds, but our generative system can evolve in any direction, spanning a vast search space for our search to navigate, potentially leading to transformational discoveries. The technical details of our implementation are described in Section~\ref{subsec_SoundGenerationSystem}.

\section{Methodology}\label{secMethod}

To facilitate evolutionary search for quality and diversity beyond the constraints of manual expert guidance, we investigate different approaches to unsupervised characterisation of sonic behaviour spaces and their dynamic redefinition. This investigation encompasses multiple strategies for feature extraction and characterisation of discovered sound objects.

Our approach employs the \ac{MAP-Elites} quality diversity algorithm to discover diverse, high-performing sound objects. \ac{MAP-Elites} operates on a structured, grid-based container where each cell stores the best solution exhibiting particular behavioural characteristics. The algorithm performs selection of occupied cells to identify previous discoveries to evolve through mutation as new elite candidates. Individuals in a new generation descending from existing container occupants are assigned a cell according to behaviour descriptors, and if the cell is either empty or the candidate achieves higher fitness according to the chosen quality metric, it is declared a new elite.

\begin{figure*}[htbp]
    \centering
    \includegraphics[width=0.85\textwidth]{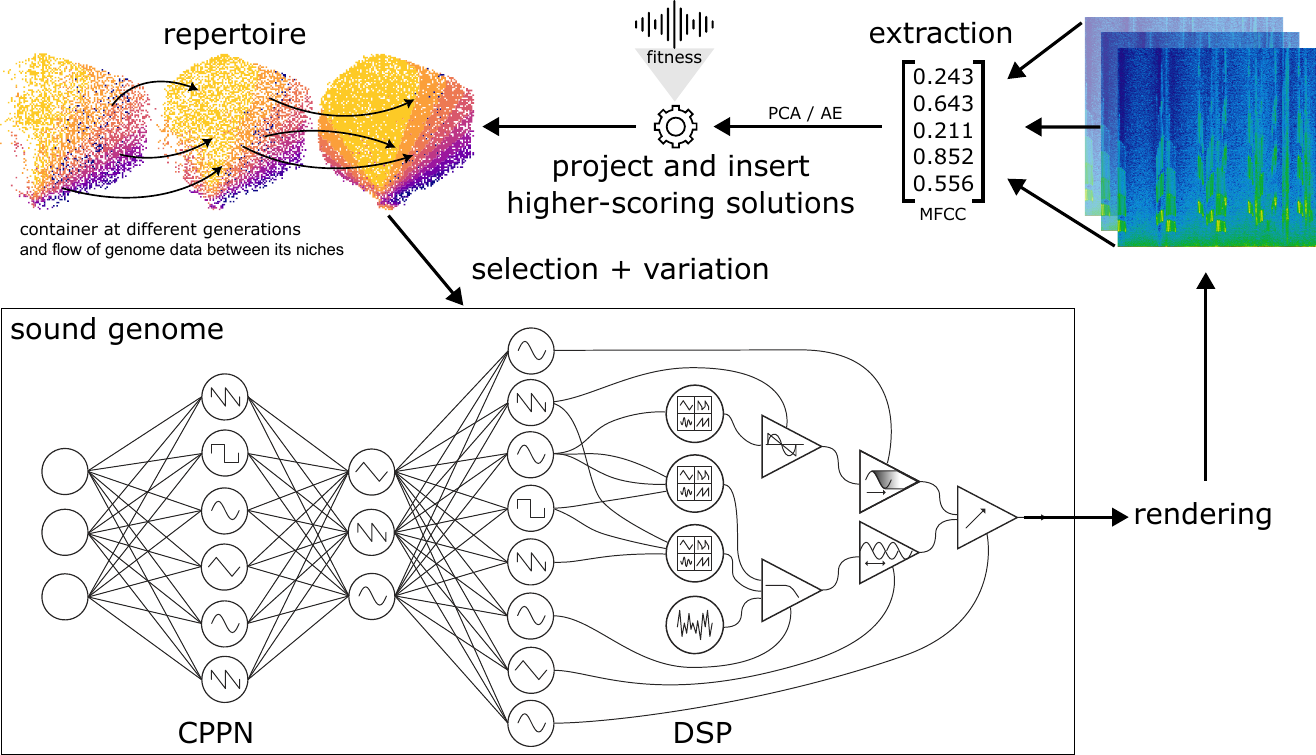}
    \caption{Visualisation of our \ac{MAP-Elites} \ac{QD} search pipeline for discovering sounds, with behaviour space container statuses at different stages of evolution on the top left, defined by unsupervised projection models.}
    \label{fig:pipeline}
\end{figure*}

Figure \ref{fig:pipeline} depicts our evolutionary \ac{QD} search pipeline. The system integrates three components: (1) sound synthesis through evolved CPPN-DSP networks, (2) feature extraction from discovered sounds, and (3) unsupervised dimensionality reduction that projects high-dimensional features onto structured behaviour spaces. We explore both static configurations, where projection models remain fixed throughout evolution, and dynamic variants that periodically retrain these models using sounds discovered during the search, allowing the behaviour space itself to adapt and evolve alongside the sound population.

We quantify both the quality of individual solutions and the diversity of the overall collection through metrics detailed in the following sections. Quality evaluation primarily employs similarity to reference sounds, though we also explore reference-free approaches based on audio problem detection (Section~\ref{subsec:MultipleReferenceExperiments} and Section~\ref{subsec:ReferenceFreeEvaluation}). Diversity is measured through pairwise distances between elite solutions in feature space.

\subsection{Sound Generation System}\label{subsec_SoundGenerationSystem}
Any sound generator, or synthesiser, is theoretically applicable to our experimental setup, as long as its configuration is evolvable, e.g. by mutation. For our experiments we use a sound synthesis architecture based on the coupling of \acp{CPPN} \cite{stanley_compositional_2007} and \ac{DSP} graphs, to both of which we apply \ac{NEAT} \cite{stanley_evolving_2002}. This method of generating sound is detailed in \cite{jonsson_system_2024} and extensions, where the CPPNs are specialised for frequency ranges, are discussed in \cite{jonsson_quality-diversity_2024}.

The CPPN networks provide audio and control signals to a DSP graph, which functions akin to a modular synthesiser patch. The CPPNs receive inputs such as a linear ramp representing time and sinusoidal signals representing pitch, allowing them to generate patterns that evolve over the duration of a sound and respond to changes in pitch. This approach enables the rendering of sounds of any duration, where the patterns encoded in the CPPNs unfold over time, revealing sub-patterns that contribute to timbral development.

Rather than using the full repertoire of mathematical functions typically available in CPPN implementations (such as Gaussian, sigmoid, or linear functions), we configure the CPPN networks to compose primarily from functions representing waveforms commonly used in modular sound synthesis: sine, square, sawtooth, and triangle 
waveforms. This choice provides a more constrained search space than arbitrary mathematical functions, potentially guiding evolution towards more sonically coherent results whilst maintaining the open-ended pattern generation capabilities of CPPNs.

We employ NEAT to evolve both the CPPN network topologies and the DSP graph structures. NEAT enables networks to start with minimal topologies and gradually complexify through the addition of nodes and connections. In our implementation, we utilise NEAT's mutation operators, 
though we do not employ its speciation mechanism or crossover operators, focusing instead on mutation-driven exploration. This allows the system to discover both interesting waveforms and effective ways to process and combine them through co-evolution of pattern generators and signal processing architectures.

A benefit of this approach is that the networks start out with the simplest possible topology and gradually complexify. By the same token, this design poses challenges as it does not encode any inductive bias \cite{locatello_challenging_2018} towards recognisable sounds, as e.g. \ac{DDSP} \cite{engel_ddsp_2019} and \ac{FM} synthesis \cite{masuda_quality-diversity_2023} do: our generative system can evolve in any direction, spanning a vast search space for our search to navigate. This can effectively require our search to undergo many iterations before approaching aesthetically pleasing or interesting sounds.

\subsection{Sonic Features}\label{subsection_Features}
There are several ways to characterize a sound by computational analysis. There are low-level acoustic features, which focus each on specific aspects, such as brightness (e.g. spectral centroid) or power spectrum mean ratios (e.g. spectral flatness). Then there are perceptual features like \acp{MFCC}, which describe the spectral shape with compact vectors. 
Here we focus on statistically extended MFCC components for multi-dimensional audio feature extraction. We also use a selection of low-level acoustic features for comparison with results obtained from evolutionary runs based on unsupervised models.
The features are used both for placement in unsupervised sonic behaviour measurement spaces and for quality evaluation.

We extract MFCC features through a multi-stage augmentation process
to capture both spectral and temporal characteristics of sounds. We begin by extracting 13 MFCC coefficients from the audio signal using frames of 400 samples with an overlap of 240 samples at a sampling rate of 16 kHz, yielding a time series of MFCC values across multiple frames for our 4-second duration sounds. We then exclude the first coefficient (MFCC0), which typically represents the overall energy of the signal, thereby focusing the feature representation on spectral shape and timbral characteristics captured by the remaining 12 MFCCs rather than overall loudness variations. 

To capture the temporal dynamics of how sounds evolve, we compute first-order derivatives (delta coefficients) for each of the 12 MFCC components. These derivatives represent the rate of change in each coefficient over time, capturing important aspects of timbral development. We then compress the temporal dimension by computing statistical summaries for both the original MFCC time series and their derivatives---specifically, the mean, standard deviation, minimum, and maximum values across all frames. This aggregation approach, following \cite{roma_general_2021}, produces 4 statistical measures for each of the 12 MFCCs (48 values) and 4 statistical measures for each of the 12 delta coefficients (48 values), resulting in a 96-dimensional feature vector that captures both spectral characteristics and temporal dynamics. This statistical aggregation provides a fixed-length representation independent of the temporal duration of the sound signal, enabling direct comparison between sounds of different lengths.

\subsection{Quality}\label{subsec_Methodology_Quality}
Previous implementations of the innovation engine algorithm in the domain of sounds have used pre-trained classifier networks \cite{jonsson_towards_2024}, which were based on similar experiments in the visual domain \cite{nguyen_understanding_2016}. In those experiments, the classifier provided definition of a behaviour space and niches within this space by the different classes it distinguishes between, and also an evaluation of quality, in the form of confidence signals for each class.

Having adopted unsupervised approaches for behaviour space definition in this work, we require alternative methods for quality evaluation that do not rely on classifier confidence signals. To this end, we investigate three approaches to quality evaluation during evolutionary sonic discovery: single-reference evaluation comparing candidates to a fixed reference sound, multiple reference evaluation comparing against collections of reference sounds, and reference-free evaluation based on audio problem detection. Each approach offers distinct trade-offs between computational efficiency, exploration bias, and perceptual quality.

\subsubsection{Single-Reference Quality Evaluation}\label{subsec:SingleReferenceMethodology}
To drive the search towards higher-quality solutions, we employ similarity metrics to arbitrarily chosen reference sounds, extracting features from target audio signals. For individual sound objects, we define quality $Q(s)$ through similarity to the chosen reference sound:
\begin{equation}
Q_{\text{single-ref}}(s) = \left(1 - d_{\text{cosine}}(\mathbf{f}(s), \mathbf{f}(r))\right)^p
\label{eq:quality_single_ref}
\end{equation}
where $\mathbf{f}(s)$ and $\mathbf{f}(r)$ represent feature vectors of the candidate and reference sounds respectively, $p$ 
is a power scaling factor, and $d_{\text{cosine}}()$ is the cosine distance metric. %
This approach maintains clear separation between quality assessment (fitness) and diversity characterisation (behaviour descriptors): whilst fitness evaluates how well individual sounds meet basic quality criteria, diversity emerges from the behaviour space structure defined by unsupervised \ac{DR}. The reference sound serves as a quality gate---a threshold for determining whether discovered sounds exhibit sufficient coherence to warrant retention---whilst leaving vast space for diverse exploration around that threshold. The \ac{MAP-Elites} algorithm maintains one elite per behaviour space cell, ensuring diversity regardless of fitness values.
In our primary configuration, we compare distances between feature vectors extracted from discovered sounds and a single sound from 
the NSynth dataset \cite{engel_neural_2017}.

For such full-reference quality assessment, we employ the cosine distance metric, where higher fitness is associated with lower distances to the given reference. This metric focuses on directional similarity and can be more robust to variations in vector magnitude, which can result in less meaningful similarity measures in high-dimensional spaces by other metrics, such as the Euclidean distance. The features extracted from discovered and reference sounds are standardised with Z-score normalisation, where the data is adjusted to present zero mean and unit variance. Global statistics for that scaling are obtained online from the evolutionary sound discovery process and the corresponding reference dataset; if e.g. only one reference sound is used from the NSynth dataset, then features from all sounds in the dataset are used for the Z-score normalisation. Features extracted from the reference datasets are available in the dataset accompanying this article \cite{jonsson_supporting_2025}.

\subsubsection{Multiple Reference Quality Evaluation}\label{subsec:MultipleReferenceMethodology}
Whilst single-reference quality assessment provides clear fitness signals by comparing all candidates to one fixed reference sound, multiple reference approaches compare discovered sounds against collections of reference sounds, providing more varied fitness signals and potentially reducing bias towards specific sonic characteristics.

We implement multiple reference evaluation using vector databases with \ac{HNSW} graphs \cite{malkov_efficient_2018} for efficient nearest neighbour search. The database is populated with features extracted from all sounds in a reference dataset (e.g., NSynth). Fitness is computed as:

\begin{equation}
Q_{\text{multi-ref}}(s) = \left(\frac{1}{k} \sum_{i=1}^{k} \left(1 - d_{\text{cosine}}(\mathbf{f}(s), \mathbf{f}(r_i))\right)\right)^p
\label{eq:hnsw_fitness_method}
\end{equation}

where $r_1, r_2, \ldots, r_k$ are the $k$ nearest reference sounds from the collection. Even when $k=1$, this approach differs 
from single-reference evaluation: whilst the single-reference approach always compares candidates to the same fixed reference sound, the multiple reference approach with $k=1$ compares each candidate to its nearest neighbour in the database, meaning the reference sound varies dynamically across different candidates. This allows the fitness landscape to adapt to different regions of the sonic space, potentially promoting more diverse exploration than a fixed reference. We experiment with $k$ values of 1 and 15 to assess the impact of neighbourhood size on diversity and quality outcomes.

\subsubsection{Reference-Free Quality Evaluation}\label{subsec:ReferenceFreeMethodology}
Obtaining measures of quality from comparison with existing sounds introduces selection bias towards specific sonic characteristics. An alternative approach evaluates quality through objective technical criteria rather than similarity to reference sounds.

We implement reference-free evaluation based on audio problem detection \cite{alonso-jimenez_automatic_2019}, 
using algorithms implemented by the Essentia library \cite{bogdanov_essentia_2013},
targeting six categories of sound artefacts: (1) clicks and discontinuities indicating synthesis instabilities, (2) gaps in audio signal suggesting synthesis failures, (3) clipping distortion from exceeded dynamic range limits, (4) noise bursts from high-energy random signals, (5) signal saturation from prolonged maximum amplitude periods, and (6) DC offset indicating improper signal centring. For each problem category $i$, we compute the proportion $P_i(s)$ as the fraction of detected events relative to the total number of analysis frames.

We supplement problem detection with compression-based quality assessment using zlib compression ratios as an indicator of signal structure and coherence. More structured, less noisy signals compress more efficiently. The compression ratio $C(s)$ is computed as:
\begin{equation}
C(s) = 1 - \frac{\text{Compressed size of } s}{\text{Original size of } s}
\label{eq:compression_ratio}
\end{equation}
where higher values indicate greater compressibility and more structured content.

The final reference-free fitness score combines all assessment components with equal weighting:
\begin{equation}
Q_{\text{ref-free}}(s) = \frac{1}{7}\left(\sum_{i=1}^{6} \left(1 - P_i(s)\right) + C(s)\right)
\label{eq:quality_ref_free_method}
\end{equation}
where $(1 - P_i(s))$ ensures that fewer detected problems result in higher quality scores.

\subsection{Unsupervised Behaviour Space Definition}
\ac{QD} algorithms rely on solution characterisations referred to as \acfp{BD}. 
A collection of such behaviours describes characteristics of discovered sound objects. Just as in robotics applications of \ac{QD} search, we face choices about how to characterise the diversity of discovered artefacts---for robotic locomotion, this might be walking distance, speed, or even flight capability; for sonic discovery, we must similarly choose how to measure and distinguish between different sound characteristics. 
We use unsupervised \ac{DR} models to automate the definition of low dimensional, structured containers for the \ac{QD} algorithm \ac{MAP-Elites}, without requiring manual selection of one combination of simple features over the other, aiming to free us from decisions based on the signal processing concepts they represent. As a reference to evaluate this approach, we compare the results from such an automatic characterization with results obtained from manually selected \acp{BD}. In both cases the containers are discretised into a $100 \times 100$ grid. 

There are a variety of approaches to consider for reducing the dimensionality of the audio features we extract from discovered sound objects.

\acf{PCA} \cite{pearson_liii_1901} provides a fast approach to constructing linear projections that capture the largest variation in the data. It produces linear combinations of input dimensions, ranked by their contribution in terms of carried information, represented by the variance. This information is obtained with linear projections to find orthogonal vectors that best fit the data. Selecting the best-fitting (principal) components is our main approach to reducing the dimensionality of the extracted audio features. 

We also experiment with non-linear approaches to automatically characterise discovered sounds by reducing their audio feature dimensionality.

Autoencoders
consist of two, jointly trained networks, for encoding and decoding \cite{rokach_autoencoders_2023}. The encoding part can be used for non-linear \ac{DR} and the decoder part can be useful for reconstruction of the original data from the lower dimensional latent space. While our experiments do not require the generative ability of the decoder, our unsupervised characterisation of sound objects may benefit from the encoder's ability to represent different levels of abstractions in the data.

\subsubsection{Behaviour Space Redefinitions}\label{subsec_Methodology_BehaviourSpaceRedefinitions}
To dynamically adapt the unsupervised, behaviour-defining DR models to the data discovered during \ac{QD} evolutionary search, 
and its evolving distribution, we periodically retrain the models at linearly increasing intervals, as discussed in \cite{grillotti_unsupervised_2022}. This approach addresses the challenge that as evolution progresses, the distribution of discovered phenotypes changes dramatically, and a behaviour space defined by the initial random population may poorly characterise the diversity present in later generations. Dynamic reconfiguration ensures that the behaviour characterisation remains relevant to the current evolutionary state whilst preventing convergence through periodic disruption of the competition dynamics.

Our approach adopts AURORA's core principle of periodic DR model retraining to enable behavioural space adaptation during evolution. However, we make different design choices that distinguish our implementation. First, whilst AURORA operates with unstructured containers and explicit size control mechanisms through adaptive distance thresholds, we investigate structured, two-dimensional grid-based containers. This maintains a fixed $100 \times 100$ grid structure that facilitates visualisation and interactive exploration whilst still allowing the behavioural characterisation to evolve. The structured container provides consistent reference points for comparison and analysis across different stages of evolution.

Second, we focus specifically on two-dimensional projections suitable for immediate visual interpretation, which could be useful in interactive and creative contexts, whereas AURORA accommodates higher-dimensional latent spaces. This constraint to 2D allows direct examination of the behaviour space and discovered sounds, though it may limit the richness of behavioural characterisation compared to higher-dimensional representations.

This combination of periodic redefinition with structured containers creates dynamic competition amongst retained solutions. As the DR models are retrained with newly discovered sounds, the relative positions of existing elites within the behaviour space shift with each reconfiguration. Solutions that were previously in different cells may be repositioned to compete within the same cell, whilst others may move to previously unoccupied regions. This dynamic repositioning maintains evolutionary pressure throughout the search and prevents stagnation in locally optimal regions of the behaviour space.

In particular, we define the generation gaps between retraining events as a triangular number sequence with increment factor 50. The generation gap $T_n$ at the $n$th retraining event is calculated using the formula:
\begin{equation}
T_n = 25n(n + 1)
\label{eq:retraining_schedule}
\end{equation}
where $n$ represents the index of the retraining event within the sequence. For example, the first five retraining events occur at generation number 50, 150, 300, 500, and 750.
As in \cite{cully_autonomous_2019}, we use the current population of elites as training data and when using an autoencoder for unsupervised \ac{BD}, we incrementally finetune the model, rather than training it from scratch, as in \cite{liapis_transforming_2013}. 

After each retraining of the projection model, the current elites in the container are re-projected to the behaviour space through the newly trained model. This can and does result in new competition dynamics, where any elite may compete with another for its original position in the behaviour space, as several elites can be projected to the same position after retraining. 
In conjunction with elite remapping, to maintain diversity and promote exploration in the search space, we employ innovation or novelty protection, such that an elite is not replaced by a new, higher-scoring competitor, which has been assigned the same position in the container after retraining, unless a protection period of 10 generations has elapsed, or if it is 10\% fitter when the protection period is in effect.

\subsection{Implementation}
The complete implementation of the methods described in this section, including the sound generation system, quality diversity search, feature extraction, quality evaluation, and unsupervised behaviour space definition, is openly available \cite{jonsson_synth-iskromosynth_2025,jonsson_synth-iskromosynth-cli_2025,jonsson_synth-iskromosynth-render_2025,jonsson_synth-iskromosynth-evaluate_2025,jonsson_synth-isphylogenetic-sequencer_2025}.

\section{Experimental Setup}\label{secExperiments}

Our 
experiments explore 
several 
variations of the evolutionary search processes,
characterized by different design choices and configurations. These configurations are defined by: how behaviours are defined, manually or automatically with unsupervised models, which are either trained initially or periodically retrained; and what types and sets of features are extracted from reference and candidate sounds, for both defining behaviours by projection to a low dimensional space and for comparing distances during quality evaluation.

At the start of each 
run, 
512 seed evaluation 
iterations 
were performed, followed by batches of 64 evaluation iterations: each batch can be defined as one generation and with a budget of 300K iterations, each simulation evaluated 4688 generations. Each experimental configuration was repeated five times to ensure statistical robustness.
Further repetitions were not performed due to the high computational cost of each run, which took approximately 48 hours on 30 \ac{HPC} CPU cores
(where genome variation, sound rendering and evaluation was parallelised between modules implemented in Node.js and Python).

We focus the experimental investigation on two core research questions through 
comparison of different evolutionary search configurations:

\textbf{Automatic vs Manual Behaviour Space Definition:} We compare automatic behaviour space definition approaches—encompassing different \ac{DR} methods (\ac{PCA} and autoencoders) and dynamic reconfiguration strategies—against manually crafted behaviour spaces using expert-selected audio features (spectral slope and rolloff - see Appendix \ref{app:DatasetCoverageInManualBD}).

\textbf{Quality Evaluation Approaches:} We assess how different quality evaluation methods affect sonic discovery effectiveness, examining reference-based quality assessment (single and multiple reference) versus reference-free evaluation methods.

To quantify the diversity of discovered solutions, we measure the average pairwise distance between all elite solutions $\mathcal{E}$ in feature space:
\begin{equation}
D(\mathcal{E}) = \frac{2}{n(n-1)} \sum_{i=1}^{n-1} \sum_{j=i+1}^{n} d_{\text{cosine}}(\mathbf{f}(s_i), \mathbf{f}(s_j))
\label{eq:diversity}
\end{equation}
where $n$ is the number of elites in $\mathcal{E}$ and $d_{\text{cosine}}()$ measures the cosine distance between feature vectors.

\section{Results}\label{secResults}

The experiments 
show 
significant differences in the ability of various approaches to discover diverse, high-quality sound objects, evaluated using the quality and diversity measures defined in Section \ref{secMethod}. All experimental data, including discovered sound objects, extracted features, and performance metrics, are available in the accompanying dataset \cite{jonsson_supporting_2025}. Sample sounds from the different experimental configurations can be auditioned online\footnote{
\url{https://video.synth.is/harpsitrees/}
}.

The quantitative results are summarised in Table~\ref{tab:comprehensive_comparison}, which presents performance metrics for the core experimental configurations addressing our research questions.

\begin{table*}[htbp]
\centering
\caption{Performance comparison across core experimental configurations.
Values represent mean ± standard deviation across 5 independent runs.
The multiple-reference and reference-free configurations were run within dynamic behaviour space definitions by periodically retrained \ac{PCA}.
}
\label{tab:comprehensive_comparison}
\begin{tabular}{lcccc}
\toprule
Configuration & Coverage & Diversity & Grid Mean Fitness & Goal Switches \\
\midrule
Dynamic PCA & 0.444 ± 0.026 & 0.514 ± 0.040 & 0.425 ± 0.014 & 34.8 ± 2.8 \\
Static PCA & 0.752 ± 0.040 & 0.512 ± 0.020 & 0.570 ± 0.011 & 7.3 ± 0.4 \\
Dynamic Autoencoder & 0.399 ± 0.079 & 0.343 ± 0.043 & 0.437 ± 0.027 & 22.9 ± 3.3 \\
Static Autoencoder & 0.636 ± 0.049 & 0.297 ± 0.050 & 0.543 ± 0.029 & 9.2 ± 0.6 \\
Manual BD & 0.886 ± 0.009 & 0.148 ± 0.004 & 0.662 ± 0.008 & 7.8 ± 0.3 \\
Multiple Ref. (k=15) & 0.497 ± 0.048 & 0.801 ± 0.049 & 0.436 ± 0.042 & 32.0 ± 3.4 \\
Multiple Ref. (k=1) & 0.482 ± 0.062 & 0.791 ± 0.023 & 0.597 ± 0.033 & 31.6 ± 2.1 \\
Reference-Free & 0.445 ± 0.055 & 0.508 ± 0.104 & 0.464 ± 0.079 & 31.9 ± 1.0 \\
\bottomrule
\end{tabular}
\end{table*}

\subsection{Automatic vs Manual Behaviour Space Definition}
\begin{figure*}[htbp]
    \centering
    \includegraphics[width=0.6\textwidth]{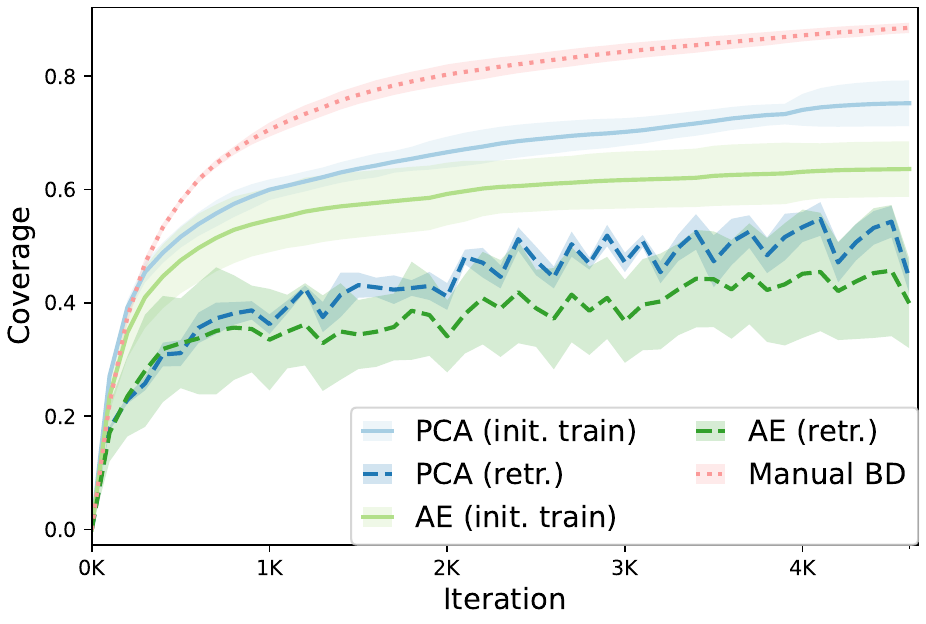}
    
    \includegraphics[width=0.6\textwidth]{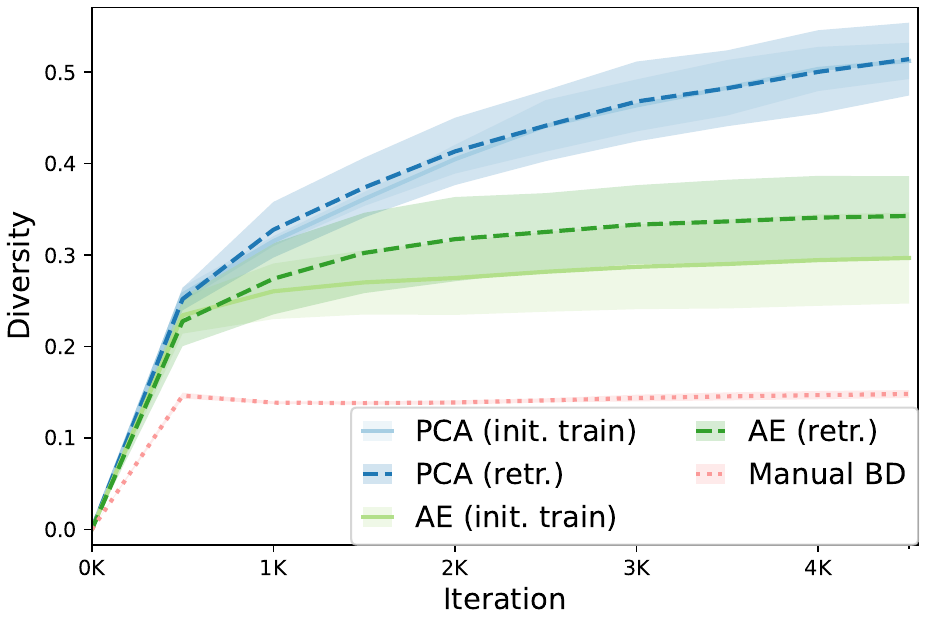}
    
    \includegraphics[width=0.6\textwidth]{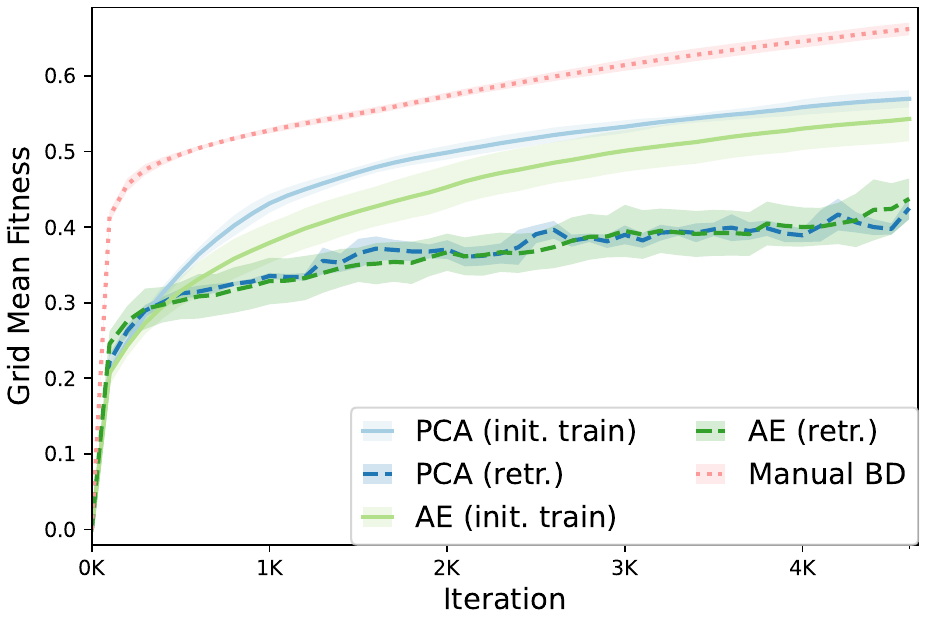}
    \caption{Performance comparison:
    Automatic vs manual behaviour space definition and dynamic vs static configuration. Solid lines show mean values across 5 independent runs. Shaded areas represent 95\% confidence intervals. Dynamic \ac{PCA} achieves significantly higher diversity than Manual BD ($p$ < 0.001), indicating that automatic approaches can outperform expert-defined spaces. \ac{PCA} shows superior diversity generation and computational efficiency, whilst autoencoders provide competitive performance with less auditory noise.}
    \label{fig:core_results}
\end{figure*}

Comparison between 
autonomous and manual 
BD configurations reveals significant advantages for 
the 
unsupervised approaches, as shown in Figure \ref{fig:core_results}. 
\ac{QD} search within \acp{BD} defined by \ac{PCA} (static and dynamic) achieves
significantly higher diversity $D(\mathcal{E})$ compared to Manual BD (0.514 ± 0.013 vs 0.148 ± 0.001, $p$ < 0.001), whilst Manual BD approaches produced results with noticeable noise components despite higher fitness scores.

In the audio domain, our approach can free experimental design from decisions based on signal processing concepts represented by individual feature descriptors. To evaluate this approach, we compare results from \ac{QD} search in automatically defined \acp{BD} with results from manually defined \acp{BD}. In previous work, such comparisons have been made where the data supplied to an unsupervised model, to automatically define a \ac{BD}, has contained all the information related to the manually defined \acp{BD} \cite{grillotti_unsupervised_2022}.

In the sound domain, where different approaches can be used to extract features, the relationship between data sources can be complex, where one source cannot be said to be strictly contained within the other. For example, when we project (extended) \ac{MFCC} features through \ac{PCA} to automatically define a \ac{BD}, a comparison with a hand-crafted \ac{BD} is not straightforward, e.g. where two instantaneous features are manually chosen to define the two axes of a behaviour space, such as spectral centroid and spectral flatness. Even though such instantaneous features and \acp{MFCC} are derived from the same underlying spectral information of the audio signal, spectral centroid and spectral flatness cannot be directly derived from \acp{MFCC}. \acp{MFCC} contain information about the overall spectral shape, which is related to but not identical to specific instantaneous features. In addition, we augment our \ac{MFCC} vectors with statistical information to capture dynamic characteristics of the sound, which the instantaneous features are less capable of representing.

Regardless of the disjoint nature of the sources of information for automatically (e.g. \ac{MFCC}) or manually defining behaviour spaces, we are curious to compare results from \ac{QD} search within those different landscapes. To that end, we rank features according to criteria based on their variance, which we take as a representation of the quantity of information they carry, and their cross-correlation (by checking if a feature provides the same information as others, down-prioritising those with a strong positive or negative correlation). Considering several feature types---such as spectral centroid, spread, skewness, kurtosis, rolloff, decrease, slope, flux, flatness and crest factor---we compute their values from audio files in the NSynth dataset ($\sim$290K files).

The highest ranking spectral features resulting from variance and correlation analysis are spectral slope, with a variance of 0.05, and spectral rolloff, with a variance of 0.04 and 0.06 max correlation with other features. Even though \ac{QD} search within such a manually defined behaviour space obtains high fitness scores, there is a noticeable noise component in the sounds and they are considerably less diverse. This indicates that unsupervised approaches may be more effective for discovering sonic diversity.

\begin{figure*}[htbp]
    \centering
    \includegraphics[width=1\textwidth]{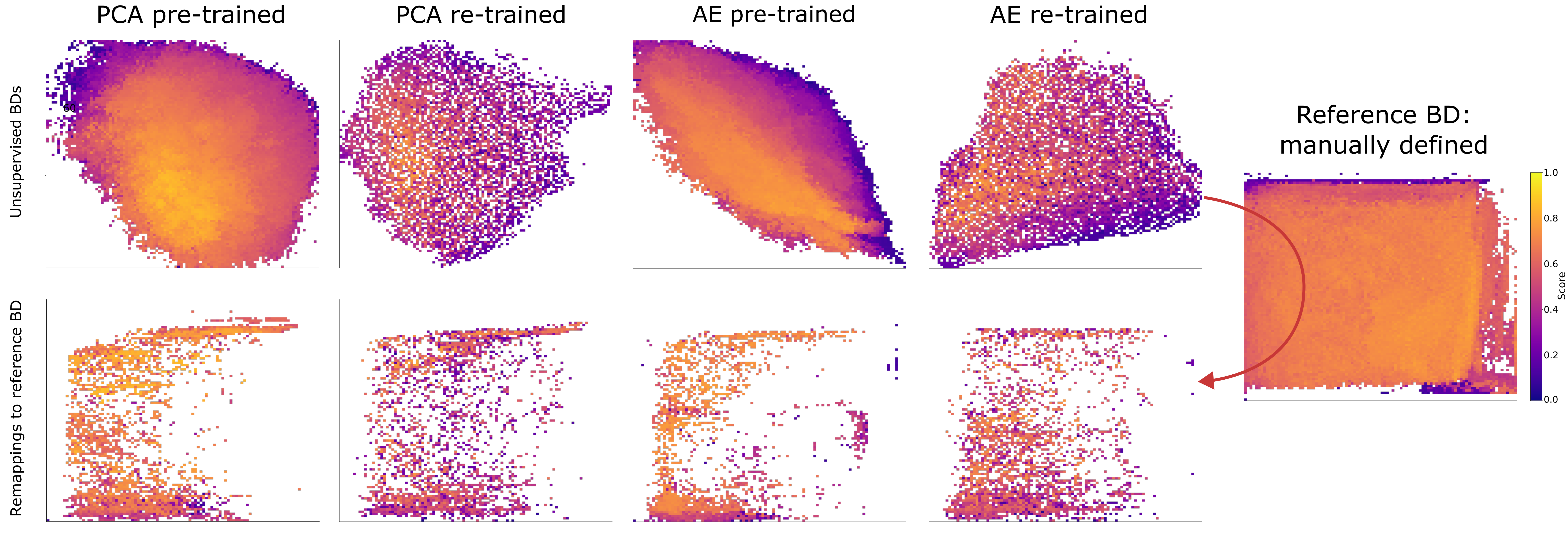}
    \caption{Coverage analysis when remapping unsupervised discoveries to manually defined behaviour space. Top row shows coverage achieved by different unsupervised approaches in their native spaces. Bottom row shows how these same discoveries map to the manual BD space (spectral slope vs rolloff). While direct search within the manual BD achieves the highest coverage (rightmost panel), remapped discoveries from automatic \acp{BD} reach most areas of the manually defined container, demonstrating that unsupervised approaches maintain substantial coverage of expert-defined sonic characteristics while achieving higher overall diversity.}
    \label{fig:remapping_analysis}
\end{figure*}

To 
compare 
different behaviour spaces directly and gain insight into how well sonic discoveries from \ac{QD} search within automatically defined search spaces address the sonic characteristics defined by our manually selected pair of audio features, we remap them into the manually defined \ac{BD}, as shown in Figure \ref{fig:remapping_analysis}. While high and continuous coverage is achieved when searching directly within the manually defined container (visible on the right of the figure), the remapped discoveries from automatic \acp{BD} do reach most areas of that specifically defined container, with results from the periodically retrained variants seemingly less constrained to specific areas. Coverage percentages for each configuration in both native and remapped spaces are provided in Table~\ref{tab:remapping_coverage}. The higher diversity measured between sonic discoveries within autonomously defined containers suggests that results from such \ac{QD} search, unconstrained by expert decisions, may also cover significant proportions of search spaces confined by measures of manually selected signal processing concepts.

\begin{table}[htbp]
\centering
\caption{Coverage percentages for different behaviour space configurations in their native spaces and when remapped to the manually defined behaviour space (spectral slope vs rolloff).}
\label{tab:remapping_coverage}
\begin{tabular}{lcc}
\toprule
Configuration & Native Coverage & Remapped Coverage \\
\midrule
PCA (Static) & 74.46\% & 22.85\% \\
PCA (Dynamic) & 41.69\% & 20.77\% \\
Autoencoder (Static) & 56.19\% & 16.46\% \\
Autoencoder (Dynamic) & 51.06\% & 21.84\% \\
Manual BD (Spectral Slope/Rolloff) & 88.18\% & --- \\
\bottomrule
\end{tabular}
\end{table}

\subsection{Dynamic Behaviour Space Reconfiguration}
Comparison between Dynamic \ac{PCA} and Static \ac{PCA} configurations reveals significant trade-offs between final performance and exploration dynamics, as demonstrated in Figure \ref{fig:core_results}. Static \ac{PCA} achieves significantly higher coverage (0.752 ± 0.013 vs 0.444 ± 0.008, $p$ < 0.001) and grid mean fitness (0.570 ± 0.004 vs 0.425 ± 0.005, $p$ < 0.001) compared to Dynamic \ac{PCA}. Regarding diversity, both configurations show similar final diversity values (Dynamic PCA: 0.514 ± 0.040 vs Static PCA: 0.512 ± 0.020, p > 0.05), indicating no statistically significant difference in final diversity between dynamic and static approaches.

However, these traditional performance metrics may not capture the full value of dynamic approaches. The periodic disruptions during retraining events create opportunities for continued exploration and prevent evolutionary stagnation, as evidenced by the sustained discovery of new elite solutions throughout the simulation rather than early convergence. While differences between using a fixed projection model and periodically retraining it produce dramatic changes in coverage and fitness, the evolutionary trajectories differ substantially, with dynamic configurations showing periodic disruptions during retraining events that may promote continued exploration and maintain evolutionary pressure, indicating an evolutionary landscape open to further innovation.

The temporary disruptions following each retraining event serve as creative destruction processes that promote continued exploration. Periodic \ac{DR} model retraining was associated with a modest diversity increase only in the autoencoder configuration, though further runs would be needed to reveal the statistical significance of this trend, and this tentative benefit comes at the cost of reduced coverage and fitness scores within the respective behaviour spaces. This suggests that the principal value of dynamic behaviour space configuration may lie less in final metric optimization and more in sustaining evolutionary potential and open-ended discovery dynamics.

Regarding the effectiveness of different \ac{DR} methods, \ac{PCA} demonstrates the most effective balance between diversity generation and computational efficiency, as shown in Figure \ref{fig:core_results}. The comparison between \ac{PCA} and autoencoders reveals 
trade-offs for sonic behaviour characterisation. \ac{PCA} achieves higher coverage and diversity values, 
indicating it is capable of broader 
exploration of sonic territories. The linear projections provided by \ac{PCA} offer computational efficiency and interpretability, allowing for rapid model retraining during dynamic configurations.

Autoencoders, whilst achieving lower overall diversity, provide notable advantages in perceptual quality. Using an autoencoder to define the search space, where the projection goes through dense layers of 64 and 32 neurons, results in competitive coverage and fitness scores. The hierarchical representations learned by autoencoders appear to better capture perceptually relevant sonic features, as evidenced by the sonic discoveries being less challenging for the listener within the more homogeneous set of discoveries. This suggests that the non-linear transformations performed by autoencoders may align more closely with human auditory perception, producing sounds that, while less diverse in feature space, are more coherent from a listening perspective.

The incremental fine-tuning capability of autoencoders during dynamic reconfiguration represents a 
methodological advantage over \ac{PCA}'s complete model replacement. This allows autoencoders to preserve learned representations while gradually adapting to evolving phenotype distributions, potentially explaining the modest diversity benefits observed in dynamic autoencoder configurations.

\begin{figure*}[htbp]
    \centering
    \begin{tabular}{@{}c@{}c@{}c@{}}
        \begin{subfigure}[t]{0.44\textwidth}
            \centering
            \includegraphics[width=\linewidth,height=10pc,keepaspectratio]{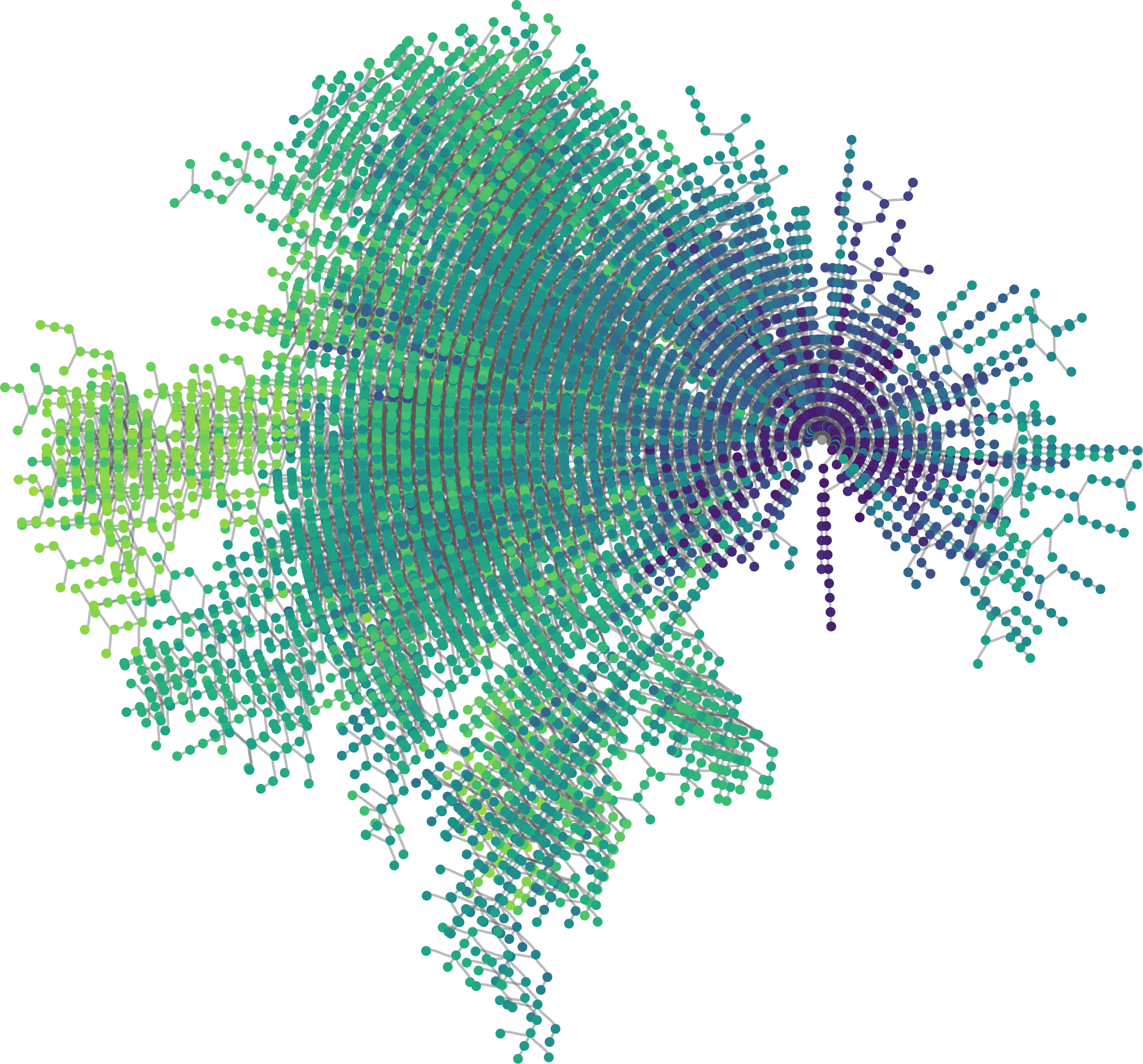}
            \caption{Static PCA (no retraining)}
            \label{fig:evolutionary_paths_A}
        \end{subfigure} &
        \begin{subfigure}[t]{0.44\textwidth}
            \centering
            \includegraphics[width=\linewidth,height=10pc,keepaspectratio]{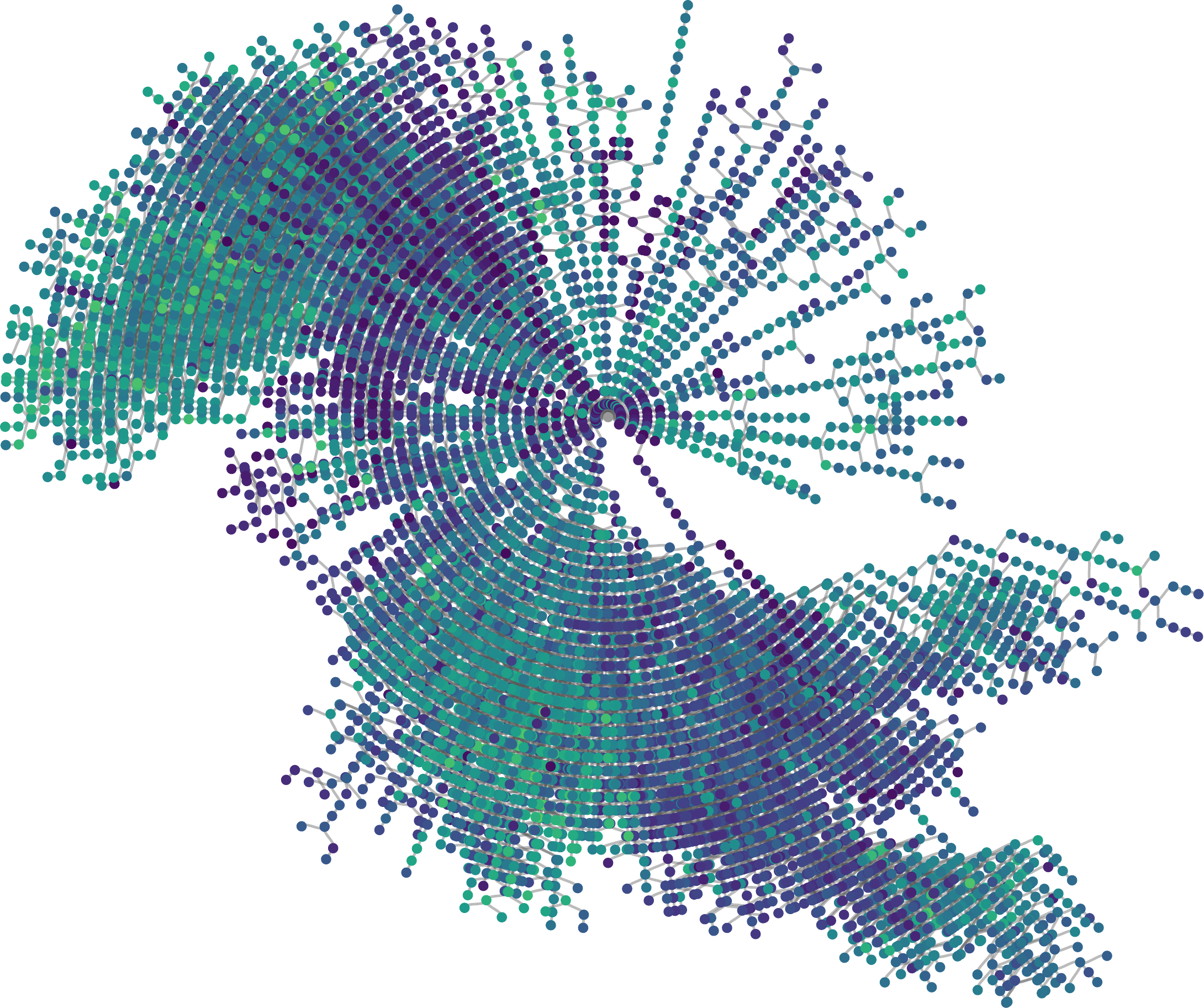}
            \caption{Dynamic PCA (periodic retraining)}
            \label{fig:evolutionary_paths_C}
        \end{subfigure} &
        \multirow{2}{*}{\raisebox{-13pc}{\includegraphics[width=0.022\textwidth,height=20pc,keepaspectratio]{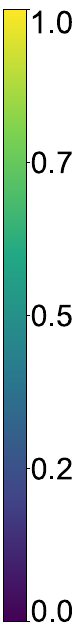}}} \\
        [3.0em]
        \begin{subfigure}[t]{0.44\textwidth}
            \centering
            \includegraphics[width=\linewidth,height=10pc,keepaspectratio]{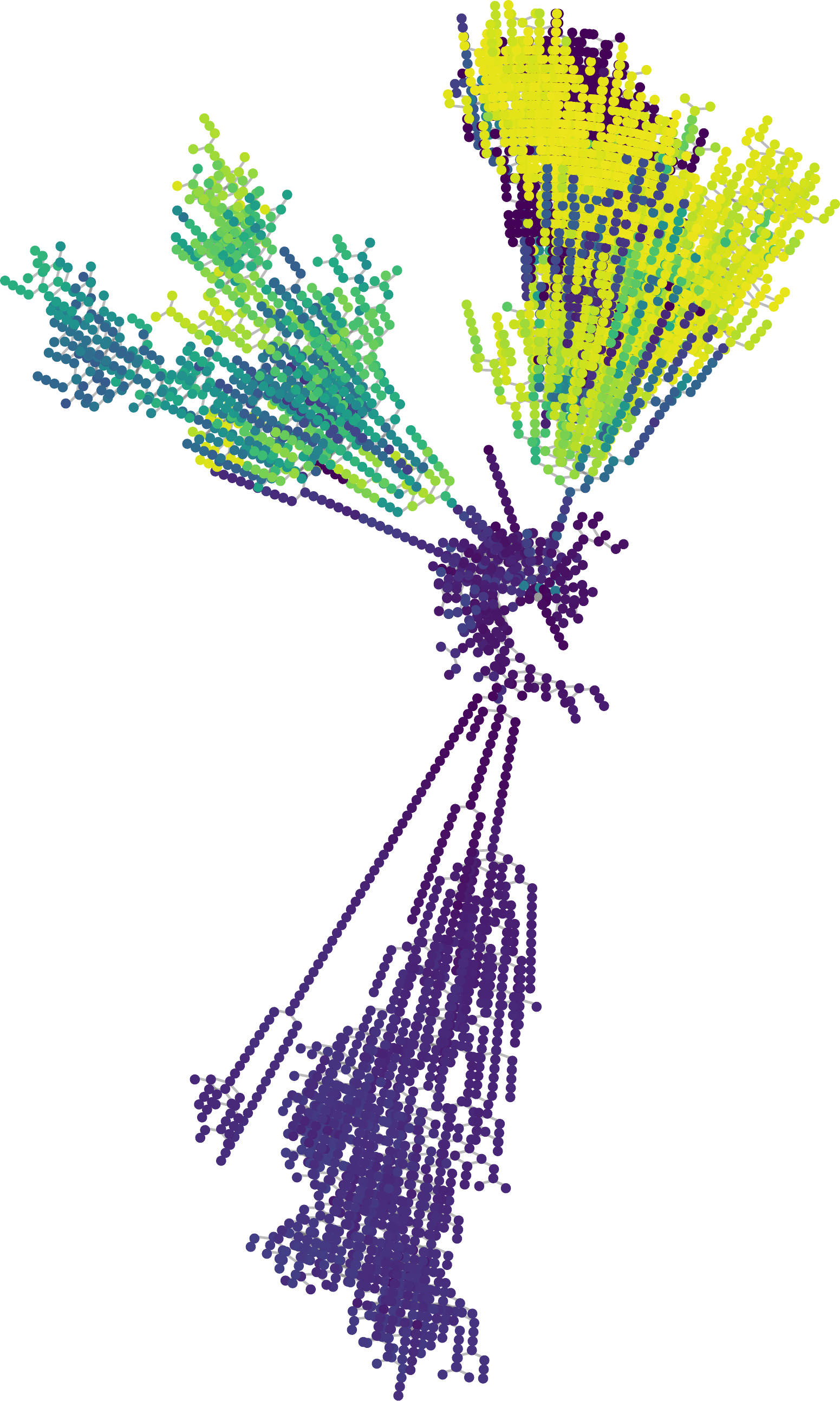}
            \caption{Manual BD (spectral slope / rolloff)}
            \label{fig:evolutionary_paths_B}
        \end{subfigure} &
        \begin{subfigure}[t]{0.44\textwidth}
            \centering
            \includegraphics[width=\linewidth,height=10pc,keepaspectratio]{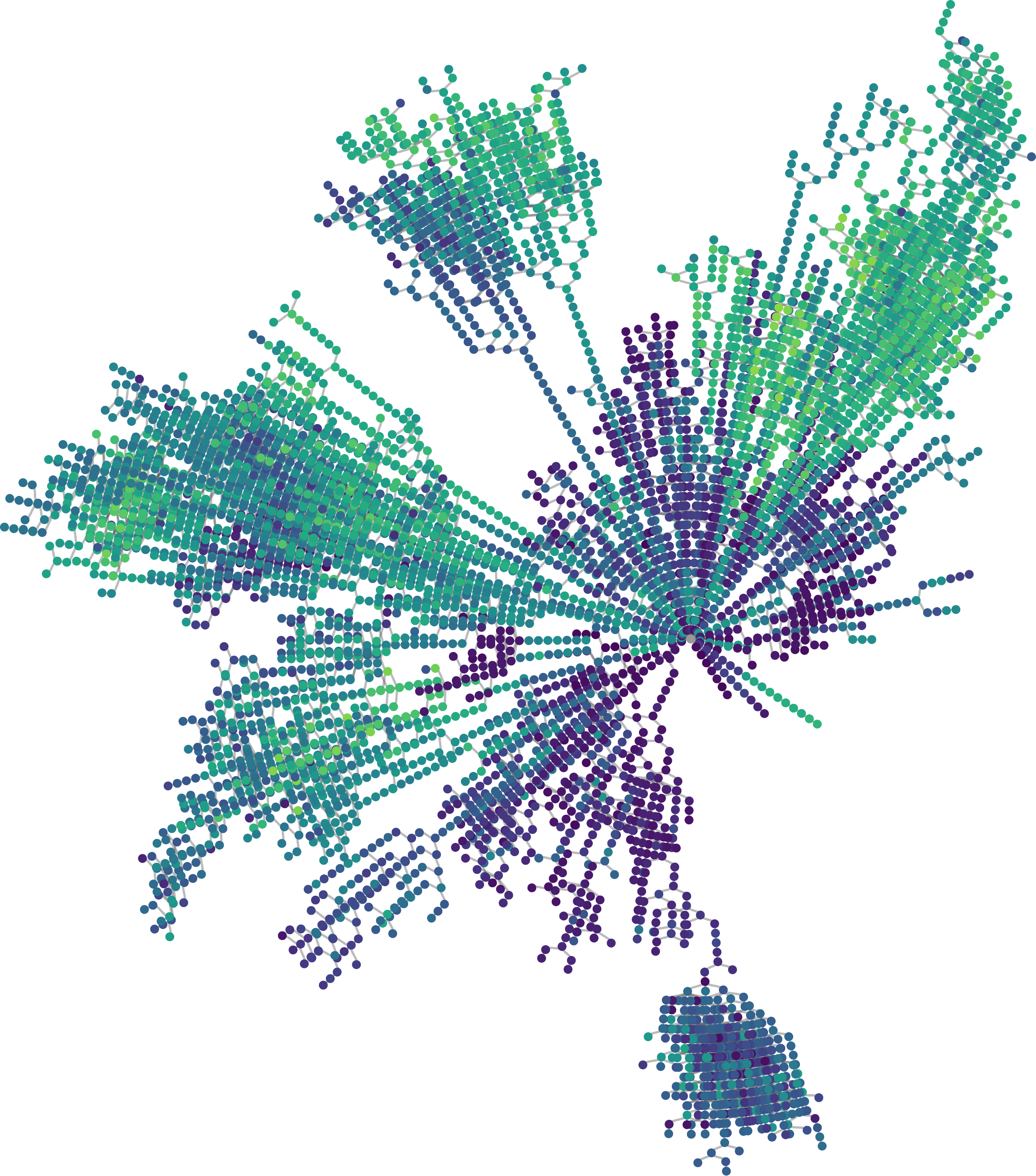}
            \caption{Dynamic Autoencoder}
            \label{fig:evolutionary_paths_D}
        \end{subfigure} & \\
    \end{tabular}
    \caption{
        Evolutionary pathways visualization showing how behaviour space definition and redefinition affect exploration patterns. 
        These radial tree diagrams display the phylogenetic tree relationships formed during the quality diversity sound search simulations, with initial seeds positioned at the centre and evolutionary branching patterns radiating outward.
        (A) Static \ac{PCA} without retraining. (B) Dynamic \ac{PCA} with periodic retraining and remapping. (C) Manually defined behaviour space (spectral slope vs rolloff).  (D) Dynamic autoencoder with incremental fine-tuning. The vertical color bar indicates the fitness used for colouring branches.}
    \label{fig:evolutionary_paths}
\end{figure*}

The evolutionary pathways reveal how dynamic reconfiguration affects exploration patterns, as visualized in Figure \ref{fig:evolutionary_paths}. Periodic retraining of the \ac{BD} model results in a higher variety of distinct evolutionary paths being more fully explored, with dynamic \ac{BD} configurations showing more diverse branching patterns compared to static approaches. The temporary disruptions following each retraining event 
serve as creative destruction processes 
that 
can help sustain
exploration.

\subsection{Quality Evaluation Approaches}
\begin{figure}[]
    \centering
    \includegraphics[width=0.6\textwidth]{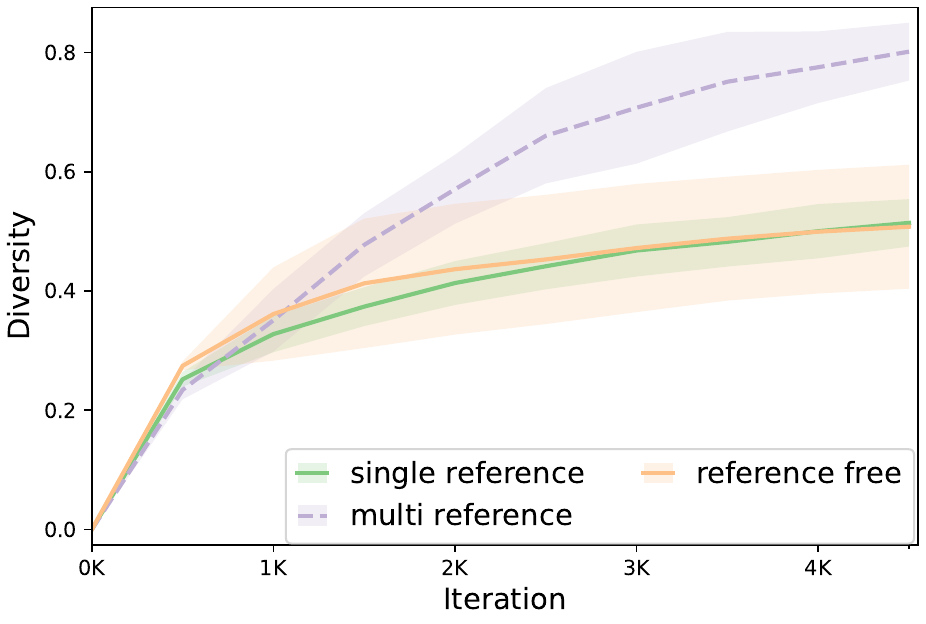}
    \caption{Performance comparison of alternative reference approaches for quality evaluation. Multiple reference approaches (with k=15) achieve the highest diversity, while reference-free evaluation provides comparable performance to single-reference methods while avoiding selection bias.}
    \label{fig:alternative_references}
\end{figure}

To address our second research question regarding different quality evaluation approaches, we 
compared 
reference-based and reference-free methods for assessing the quality of discovered sounds during evolutionary search.
Figure~\ref{fig:alternative_references} demonstrates the comparative performance across different quality evaluation approaches.

\subsubsection{Multiple Reference Quality Evaluation}\label{subsec:MultipleReferenceExperiments}

We evaluated the impact of multiple reference quality assessment on sonic discovery effectiveness. Experimental comparison demonstrates that multiple reference approaches achieve higher diversity than single reference methods. The $k=15$ configuration achieved diversity of 0.801 ± 0.049 
and dynamic single reference with $k=1$ resulted in similar diversity of 0.791 ± 0.023.
Comparing this to the diversity of 0.514 ± 0.040 achieved by runs using a fixed, single reference for quality evaluation (both within dynamic \ac{PCA} \acp{BD})
indicates that exposing the evolutionary process to multiple reference targets can promote exploration of diverse sonic territories whilst maintaining quality standards.

Coverage metrics showed comparable performance across configurations (Multiple Ref k=15: 0.497 ± 0.048; Multiple Ref k=1: 0.482 ± 0.062), whilst grid mean fitness values demonstrated variation (k=15: 0.436 ± 0.042; k=1: 0.597 ± 0.033). The higher fitness with single nearest neighbour comparison reflects more focused selection pressure, whilst the broader neighbourhood averaging promotes diverse exploration. Both multiple reference configurations maintained high goal switching frequencies (k=15: 32.0 ± 3.4; k=1: 31.6 ± 2.1), indicating sustained exploration throughout evolution comparable to dynamic behaviour space configurations.

\subsubsection{Reference-Free Quality Evaluation}\label{subsec:ReferenceFreeEvaluation}

Reference-free evaluation methods (within dynamic \ac{PCA} \acp{BD}) demonstrate that quality assessment without reference sounds can achieve comparable diversity whilst avoiding selection bias towards specific sonic characteristics. Reference-free evaluation achieved diversity of 0.508 ± 0.104, comparable to single reference approaches and indicating that technical quality criteria can guide discovery of diverse sonic territories without requiring similarity to existing sounds.

Sounds deemed high-quality by reference-free metrics showed varied characteristics and could potentially be useful in creative applications. The absence of reference comparison bias allowed exploration of sonic characteristics that might otherwise be under-represented when fitness is determined through similarity measures. Coverage (0.445 ± 0.055) and grid mean fitness (0.464 ± 0.079) values fell within the range of other experimental configurations. Goal switching frequency (31.9 ± 1.0) matched dynamic behaviour space configurations, indicating sustained evolutionary exploration throughout the simulation.

\subsection{
Goal Switches
}

\begin{figure*}[htbp]
    \centering
    \begin{tabular}{@{}c@{}c@{}}
        \multicolumn{2}{@{}c@{}}{\includegraphics[width=1.0\textwidth]{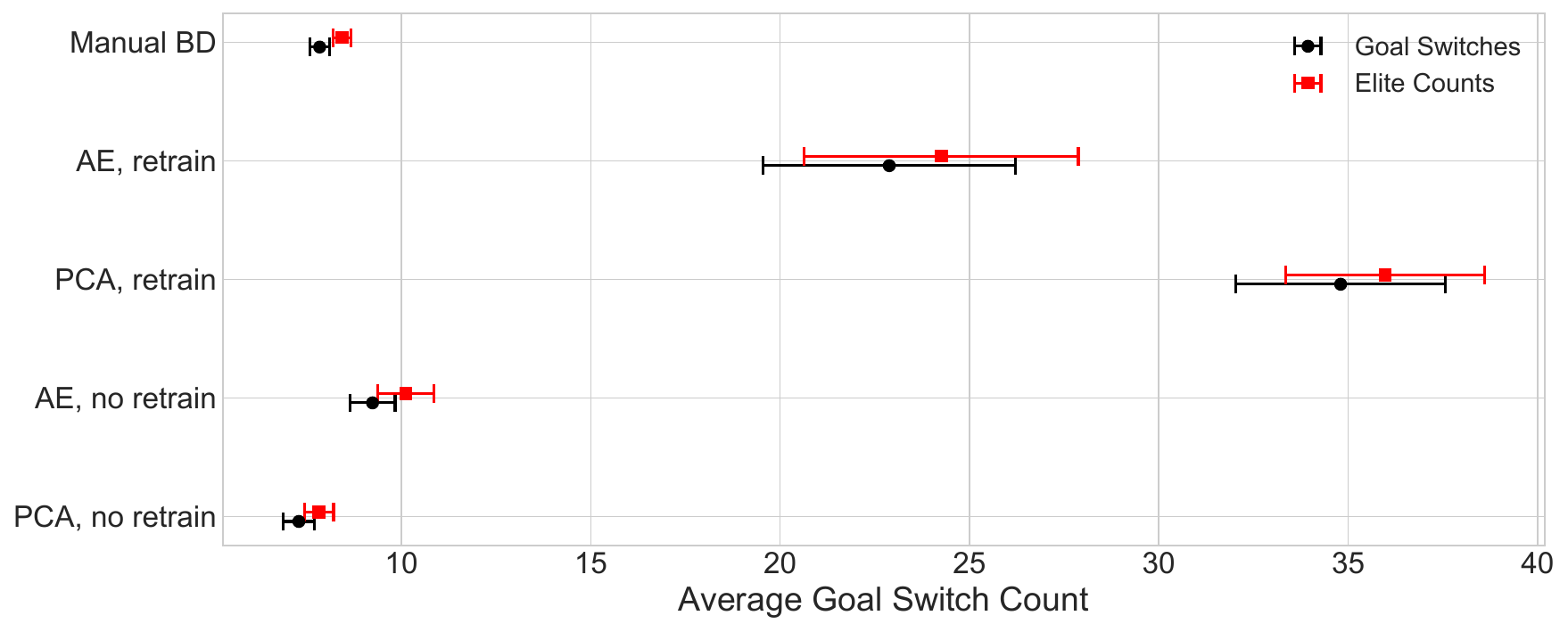}} \\
    \end{tabular}
    \caption{
    Niche-centric goal switching showing average elite counts (red squares) and cross-niche descendants (black circles). 
    }
    \label{fig:goal_switches}
\end{figure*}

A measure of \textit{goal switches} can give insights into how the diversity of possible behaviours in the search space has been leveraged during the evolutionary process. Another way to interpret this metric is how often stepping stones, between different niches in the search space, have been collected. This measure has been defined 
in a niche-centric manner 
as the number of times during a run that a new elite in one cell was the offspring of an elite of another cell (\cite{nguyen_innovation_2015,nguyen_understanding_2016}).
The forest plot in figure \ref{fig:goal_switches} shows the 
goal switches 
during the different variants of \ac{QD} search, from the perspective of individual cells: the square (red) dots show how often on average a new elite was crowned in the populated cells, and the circular (black) dots show how often on average a new occupant was a descendant of an elite from another cell, which can be interpreted as one pair of stepping stones.

When not periodically retraining the \ac{DR} model, such stepping stone collection is much less frequent than when the \ac{BD} is defined by dynamically configured, unsupervised projections. The goal switches occurring in variants based on dynamic \ac{DR} models are more frequent than the 21.7\textpm3.6 reported from sonic innovation engine experiments based on a pre-trained \ac{DNN} classifier \cite{jonsson_quality-diversity_2024}. This can be partly attributed to the population reprojection after each \ac{DR} model retraining event, where each member can be assigned to a new niche, and declared its occupant if it performs best among other competitors assigned to the same niche after redefinition of the search landscape.

It is interesting to observe in Figure \ref{fig:goal_switches} that new cell elites come almost always from different cells---by how close goal switches (black circles) are to average new elite counts (red squares)---while in those previous studies that was only the case in 63.2\% of new elite settlements. This seems to indicate that the evolutionary variants under study here are more aggressively collecting stepping stones across the 
sonic search space.

\section{Discussion}\label{secDiscussion}

Signal processing features selected for technical analysis may not fully capture the kinds of sonic distinctions that emerge during open-ended exploration. Our results show notable differences between automatic and manual approaches to characterising sonic diversity, with automatic behaviour space definition through unsupervised dimensionality reduction outperforming expert-selected features in discovering diverse sounds, though manually defined spaces achieved higher coverage and fitness scores.

\subsection{Automatic vs Manual Behaviour Space Definition} Our comparison of automated and expert-defined behaviour characterisation revealed significant advantages for unsupervised approaches, as shown in Figure~\ref{fig:core_results}. Automatic behaviour space definition through Static and Dynamic \ac{PCA} achieved significantly higher diversity than manually crafted behaviour spaces using spectral slope and rolloff, indicating that unsupervised methods can outperform expert-selected features for sonic discovery. This advantage persisted despite manually defined spaces achieving higher fitness scores and coverage, suggesting a 
disconnect between signal-processing-motivated features and sonic diversity. The resulting sounds from manual behaviour spaces contained noticeable noise components despite their higher quantitative performance. This suggests that expert-selected features like spectral slope and rolloff, while well-motivated from an audio signal-processing perspective, may not fully capture the sonic qualities that lead to diverse discoveries. The remapping analysis revealed that automatic discoveries still cover substantial portions of expert-defined characteristic spaces, indicating that automatic approaches maintain breadth whilst producing sounds with less noticeable noise artefacts.

Regarding the effectiveness of different automatic approaches, \ac{PCA} proved the most effective \ac{DR} method, balancing diversity generation with computational efficiency. Autoencoders provided competitive performance with less auditory noise, suggesting that their hierarchical representations may better capture certain sonic features. The incremental fine-tuning capability of autoencoders during dynamic reconfiguration represents a methodological advantage over \ac{PCA}'s complete model replacement. When retraining occurs, \ac{PCA} must be recomputed entirely from the current elite population, discarding all previous model parameters and potentially causing abrupt shifts in behaviour space structure. In contrast, autoencoders can be incrementally fine-tuned from their previous weights, allowing the network to gradually adjust its learned representations while maintaining continuity with earlier projections. This preservation of learned structure may reduce the disruptive effects of remapping events, potentially explaining the modest diversity benefits observed in dynamic autoencoder configurations.

The relationship between dynamic reconfiguration and performance proves more complex than initially anticipated, as shown in Figure~\ref{fig:core_results}. Static \ac{PCA} achieved superior coverage and grid mean fitness compared to Dynamic \ac{PCA}, yet both configurations showed similar final diversity. However, the evolutionary trajectories differ: 
dynamic configurations demonstrated sustained exploration throughout the simulation rather than early convergence, with goal switching frequencies far exceeding static approaches. This suggests that the value of dynamic reconfiguration lies not in final metric optimisation but in maintaining evolutionary potential and preventing stagnation.

\subsection{Quality Evaluation Approaches} The investigation of different quality evaluation methods shows varied impacts on sonic discovery effectiveness. Reference-based approaches using single targets provide clear fitness signals but may introduce bias toward specific sonic characteristics. Multiple reference approaches, comparing discovered sounds against collections rather than individual targets, achieved higher diversity compared to single reference methods, as shown in Figure~\ref{fig:alternative_references}, indicating that broader target exposure can promote more diverse exploration.

An important distinction exists between single-reference and multiple reference approaches that extends beyond neighbourhood size. Even when $k=1$, the multiple reference approach 
differs from single-reference evaluation: whilst the single-reference approach creates a fixed fitness landscape by comparing all candidates to the same reference sound, the multiple reference approach with $k=1$ creates a dynamic, adaptive fitness landscape where each candidate is compared to its nearest neighbour in the reference database. This dynamic reference assignment allows different regions of the sonic space to be evaluated against locally appropriate references, potentially explaining the higher diversity achieved by multiple reference approaches. The fitness landscape effectively adapts to the candidate's location in feature space, providing contextually relevant quality assessment rather than a global standard. This spatial adaptation may reduce the convergence pressure towards a single sonic archetype whilst maintaining meaningful quality constraints.

Reference-free evaluation methods based on audio problem detection achieved comparable diversity to single reference approaches whilst avoiding selection bias, indicating viable alternatives for unbiased quality assessment. The choice of quality evaluation approach represents a significant factor in balancing exploration breadth with target coherence in evolutionary sonic discovery.

\subsection{Broader Implications}

Our \ac{QD} approach employs fitness for quality assessment within behavioural niches rather than global optimisation. Quality constraints ensure discovered sounds meet minimum coherence criteria, whilst behaviour space structure drives exploration of diverse sonic territories. The results indicate how challenging it can be to manually identify audio features that lead to diverse discoveries. When favouring exploration over exploitation, it is difficult to anticipate what audio features, high and low level, may reveal audio characteristics that prove inspiring to a sound artist.

Though unconstrained, the generative system employed in this and related works \cite{jonsson_system_2024,jonsson_towards_2024} has its limitations. Those can be viewed in light of \ac{QD} experiments with robotic locomotion, where agents (usually) don't invent flight. Similarly, the sound generator employed in our experiments can't invent all possible timbres. Future work could extend these investigations to other generative systems and experimental setups.

The influence of periodic retraining has been varied, depending on the type of model, with autoencoders leading to a higher diversity of sound discoveries when incrementally retrained, while the diversity discovered within behaviour spaces defined by \ac{PCA} seems not influenced by such periodic re-definition based on the current population of elites. Retraining the models during \ac{QD} search results in lower coverage and score within their respective behaviour spaces, while results from those configurations show different spatial distribution patterns when remapped to the manually defined search space, with dynamic approaches reaching different corners of the behaviour space despite slightly lower overall coverage, as visualized in Figure \ref{fig:remapping_analysis}.

Our findings regarding the limitations of manually defined behaviour spaces provide insight into why different \ac{QD} approaches suit different sonic applications. Masuda \& Saito's success with expert-selected spectral features for sound matching \cite{masuda_quality-diversity_2023} makes sense when the goal is finding diverse ways to approximate known targets. However, our results indicate that such manual definitions can significantly limit diversity in open-ended discovery contexts, where the goal is exploring unknown sonic territories rather than matching familiar ones.

\section{Conclusion}\label{secConclusion}

We investigated the application of \ac{DR} techniques to automatically define and dynamically reconfigure behaviour spaces for evolutionary sound discovery. The experimental design combined periodic retraining of DR models to adapt behaviour characterisation, a generative system (CPPN-DSP networks) that can evolve both structure and parameters without architectural constraints, and quality evaluation independent of supervised classifiers. This prioritised exploration over exploitation of expert knowledge. Through experimental comparison, we have shown the viability of unsupervised sonic landscape illumination that can overcome limitations of expert-defined search spaces.

The investigation reveals that unsupervised approaches 
outperform manual behaviour space definition in discovering sonic diversity, despite achieving lower coverage in expert-defined feature spaces. This finding indicates limitations in relying solely on signal processing intuitions for creative exploration and suggests that meaningful sonic diversity may emerge from relationships not captured by conventional audio analysis.

In addition, we explored different quality evaluation approaches. Multiple reference evaluation methods achieved higher diversity compared to single-reference approaches, whilst reference-free evaluation based on audio problem detection provided comparable diversity without selection bias toward specific sonic characteristics. The choice of quality evaluation approach represents an important factor in balancing exploration breadth with target coherence.

Dynamic behaviour space reconfiguration presents a nuanced trade-off between final performance metrics and evolutionary dynamics. Whilst static approaches achieve superior coverage and fitness scores, dynamic reconfiguration helps maintain exploration potential throughout evolution, reducing the risk of stagnation that can accompany convergence to local optima. These results indicate that considering evolutionary dynamics alongside final metrics may be important for creative applications.

Among \ac{DR} approaches, \ac{PCA} provides the most effective balance for sonic applications, though autoencoders offer advantages in producing less auditory noise. The computational simplicity and interpretability of \ac{PCA} offer practical advantages, whilst the hierarchical representations learned by autoencoders may better capture certain sonic relationships.

These findings 
indicate 
that meaningful creative exploration can emerge from computational processes without requiring continuous human aesthetic guidance or domain expertise. The ability to automatically discover sonic distinctions represents a step toward systems capable of more autonomous sonic exploration.

Future investigations could explore integration of novelty archives with dynamic behaviour spaces, application to higher-dimensional unstructured containers, and extension to alternative sound generation architectures. Human perceptual evaluation would strengthen validation of computational diversity measures, while real-time applications could explore the creative potential of these approaches for interactive music systems.

\section*{Author contributions}
All authors contributed to the study conception and design. Simulation design, execution, analysis and corresponding software development were performed by Björn Þór Jónsson. The first draft of the manuscript was written by Björn Þór Jónsson with review input and corrections from all authors.

\section*{Acknowledgments}
This work was performed on Educloud Fox – High Performance Computing cluster, owned by the University of Oslo IT Department. Additional computations were performed on the Norwegian Research and Education Cloud (NREC), using resources provided by the University of Bergen and the University of Oslo. Data storage and analysis was performed on resources provided by Sigma2 - the National Infrastructure for High-Performance Computing and Data Storage in Norway, project 9648K. 

\section*{Financial disclosure}
This work was supported by the Research Council of Norway through its Centres of Excellence scheme, project number 262762.

\section*{Conflict of interest}
The authors declare no potential conflict of interests.

\bibliography{references-BibTeX}

\appendix

\section{Dataset Coverage in Manually Defined Behaviour Spaces}\label{app:DatasetCoverageInManualBD}

\begin{figure*}[htbp]
    \centering    
    \begin{tabular}{@{}c@{\hspace{1em}}c@{}}
        \includegraphics[width=0.48\textwidth,height=0.22\textheight,keepaspectratio]{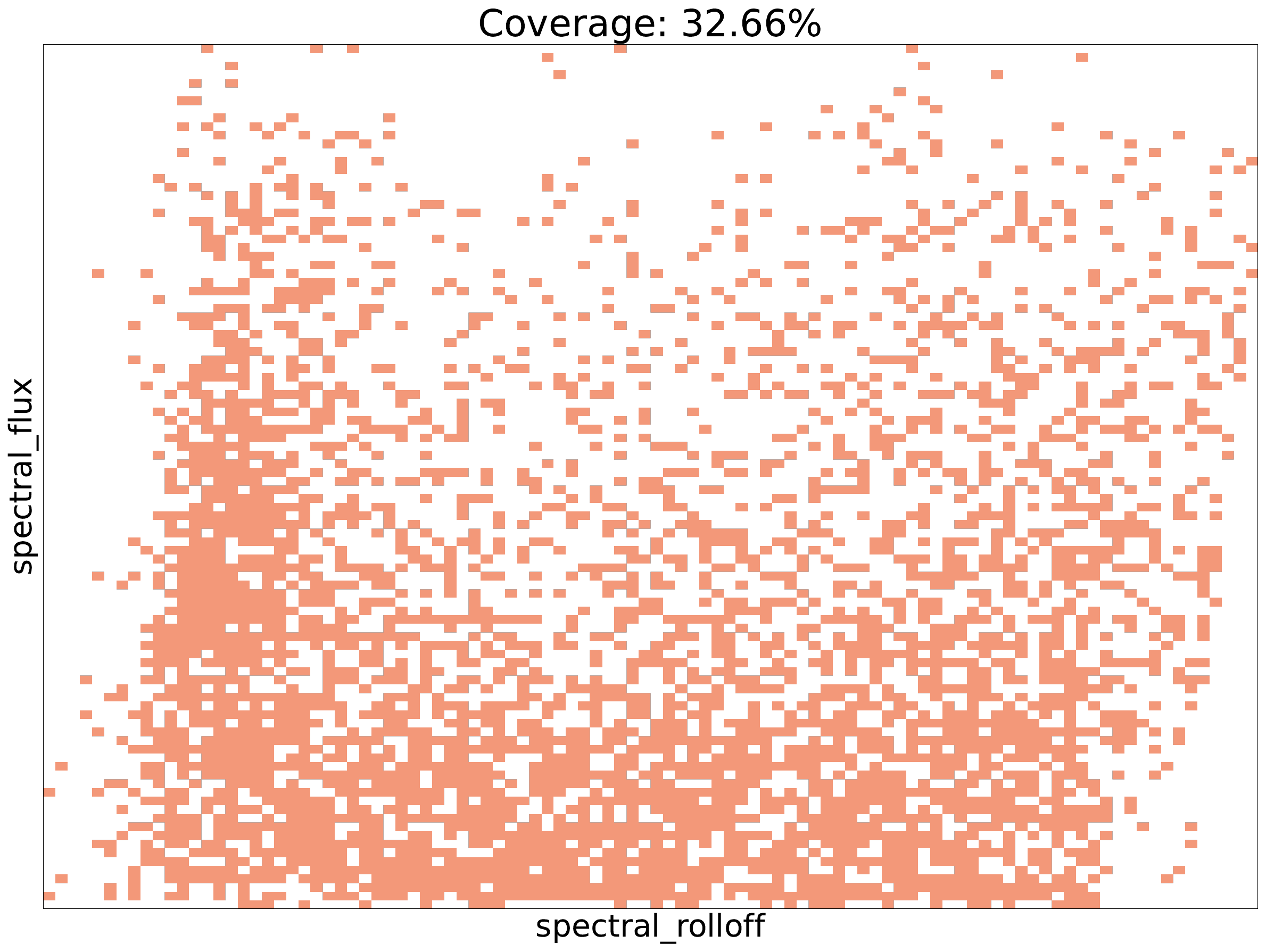} & \includegraphics[width=0.48\textwidth,height=0.22\textheight,keepaspectratio]{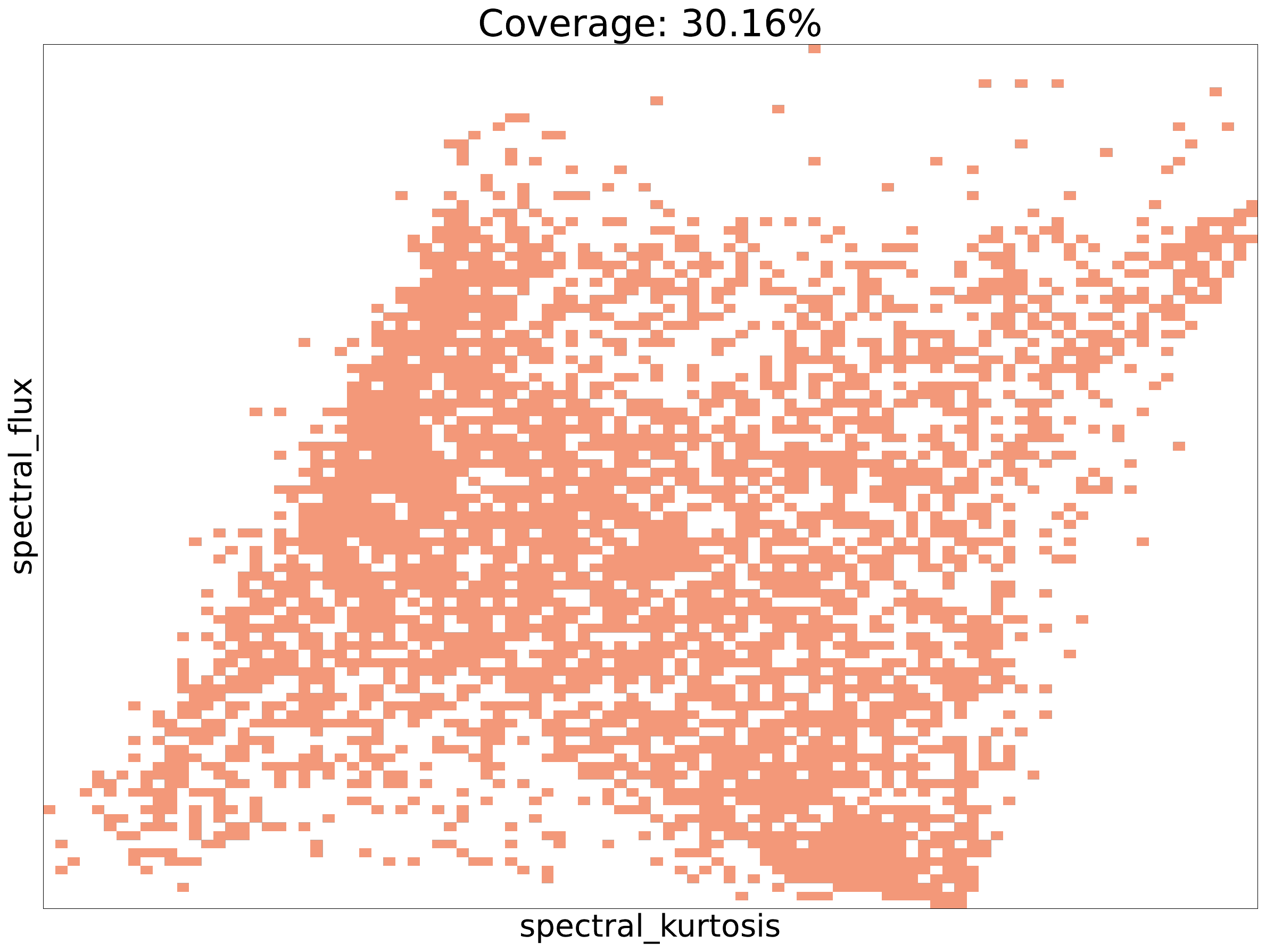} \\[1em]
        \includegraphics[width=0.48\textwidth,height=0.22\textheight,keepaspectratio]{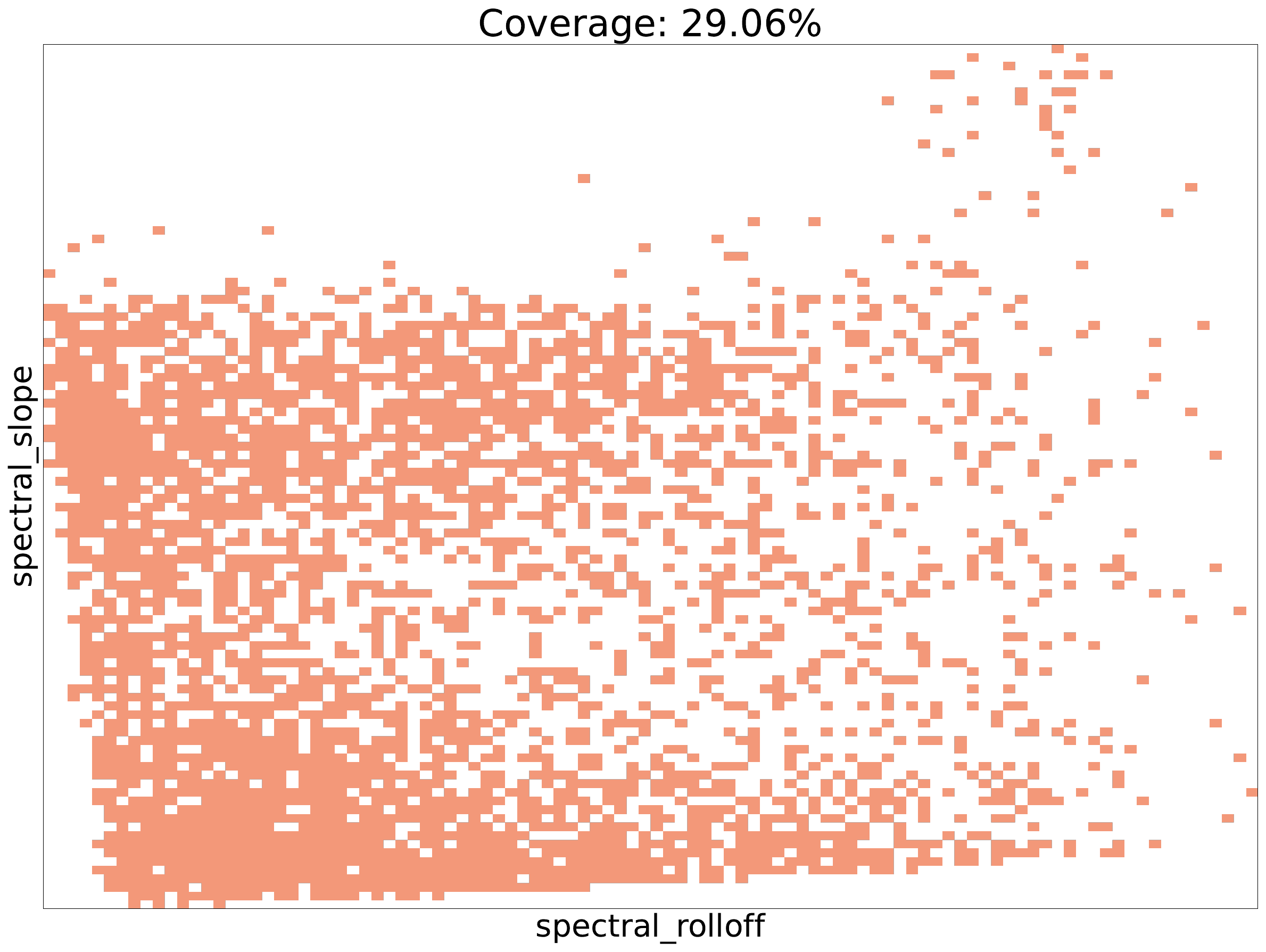} & \includegraphics[width=0.48\textwidth,height=0.22\textheight,keepaspectratio]{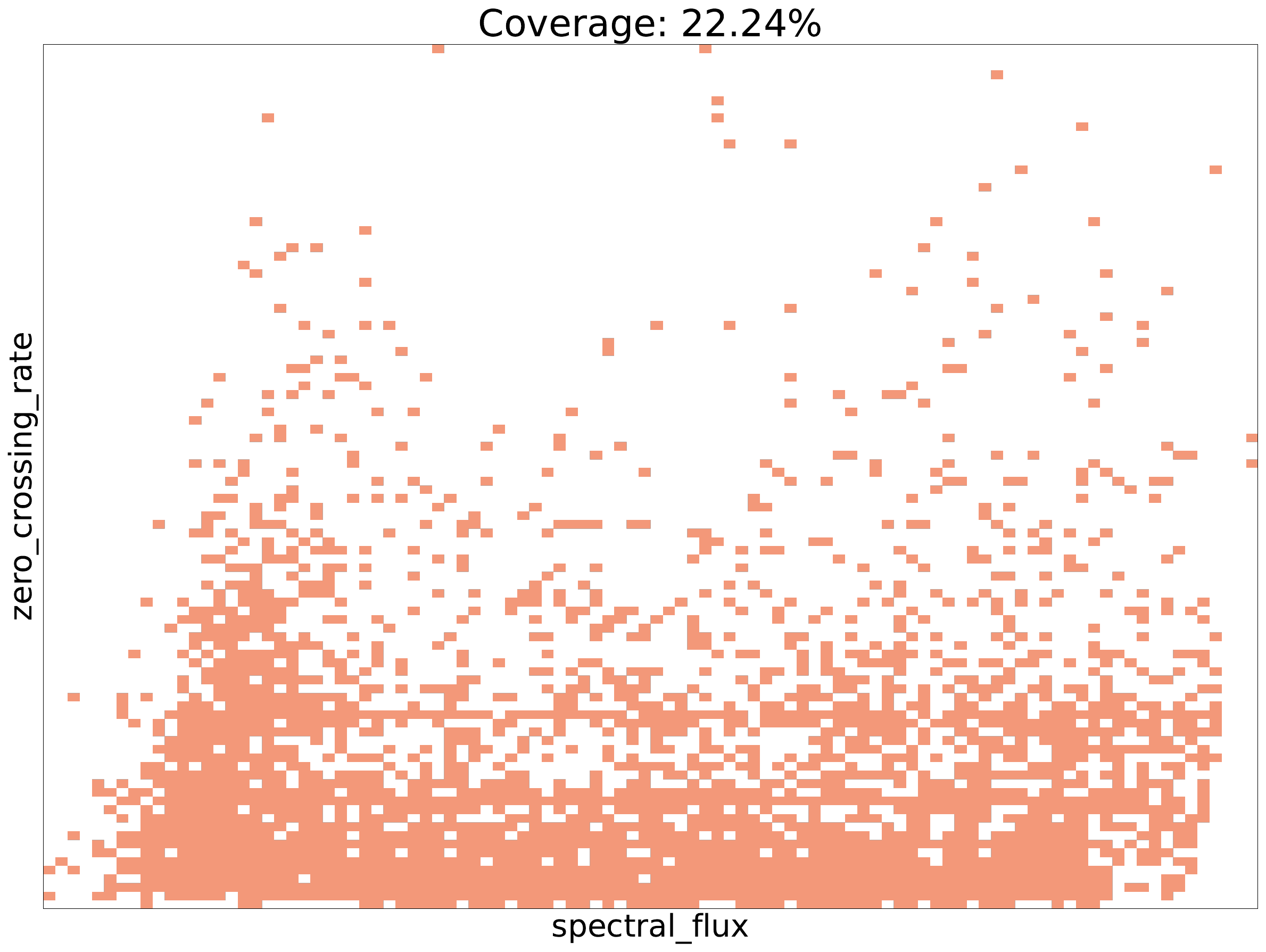} \\[1em]
        \includegraphics[width=0.48\textwidth,height=0.22\textheight,keepaspectratio]{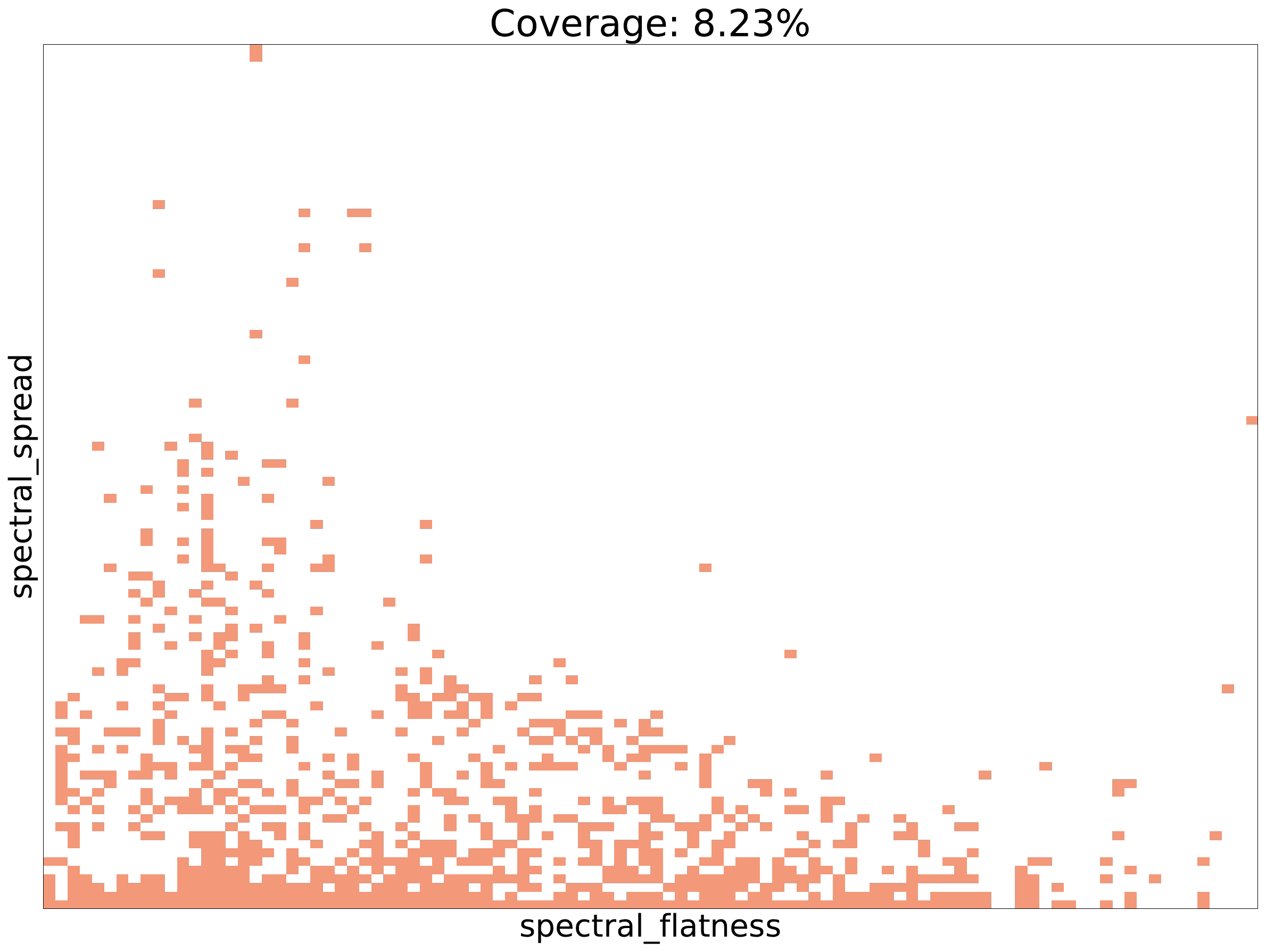} & \includegraphics[width=0.48\textwidth,height=0.22\textheight,keepaspectratio]{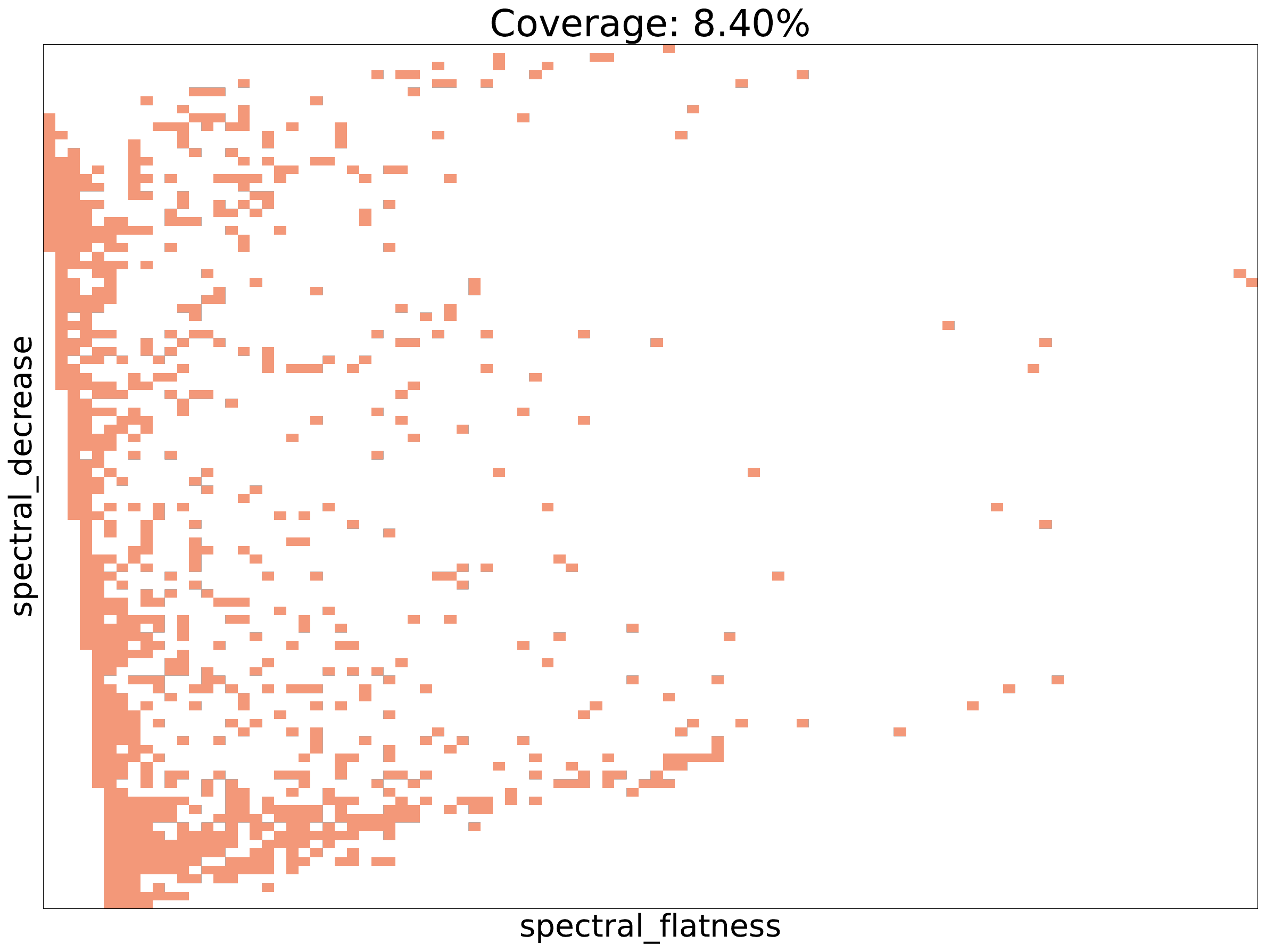} \\[1em]
        \includegraphics[width=0.48\textwidth,height=0.22\textheight,keepaspectratio]{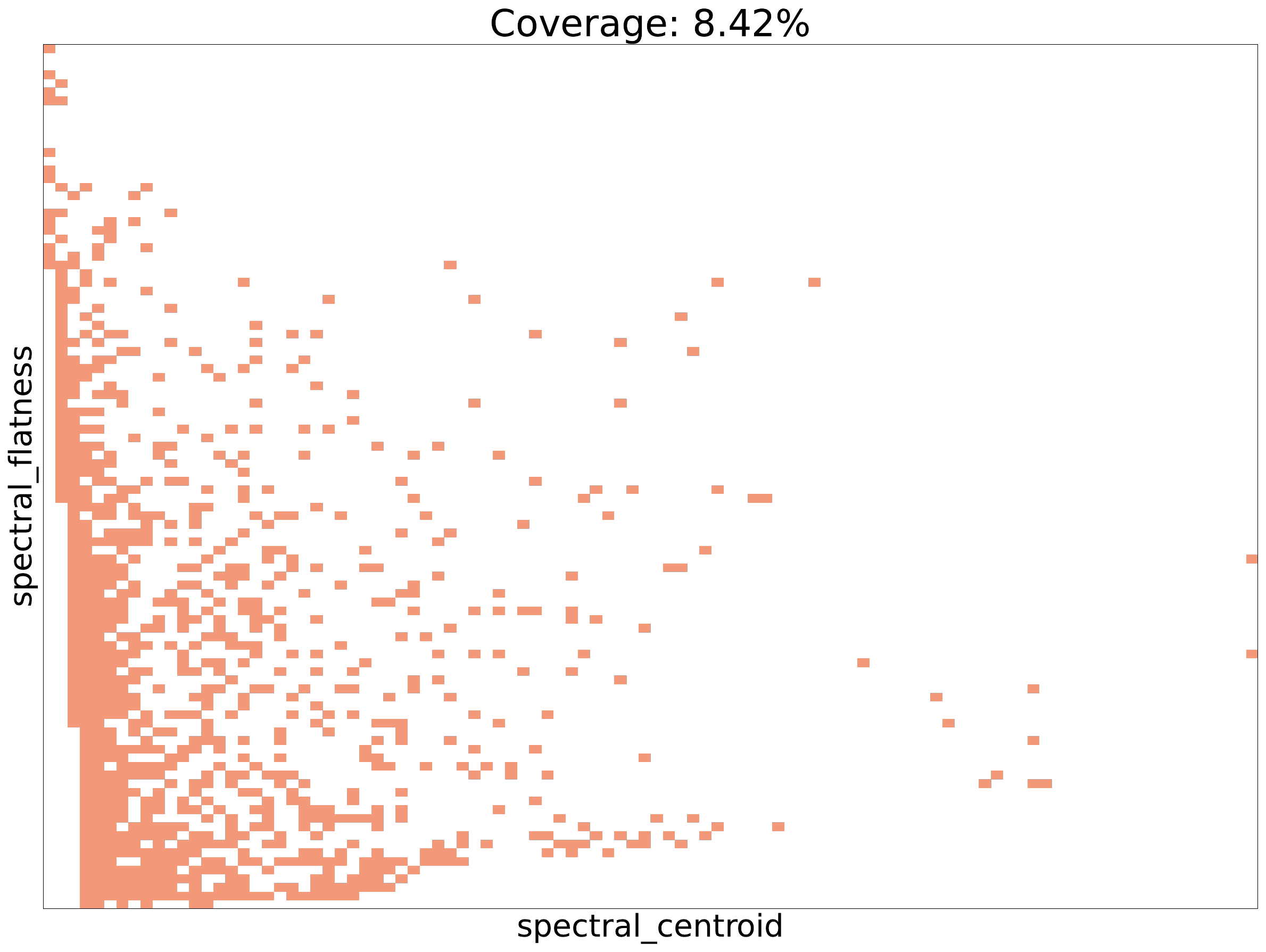} & \includegraphics[width=0.48\textwidth,height=0.22\textheight,keepaspectratio]{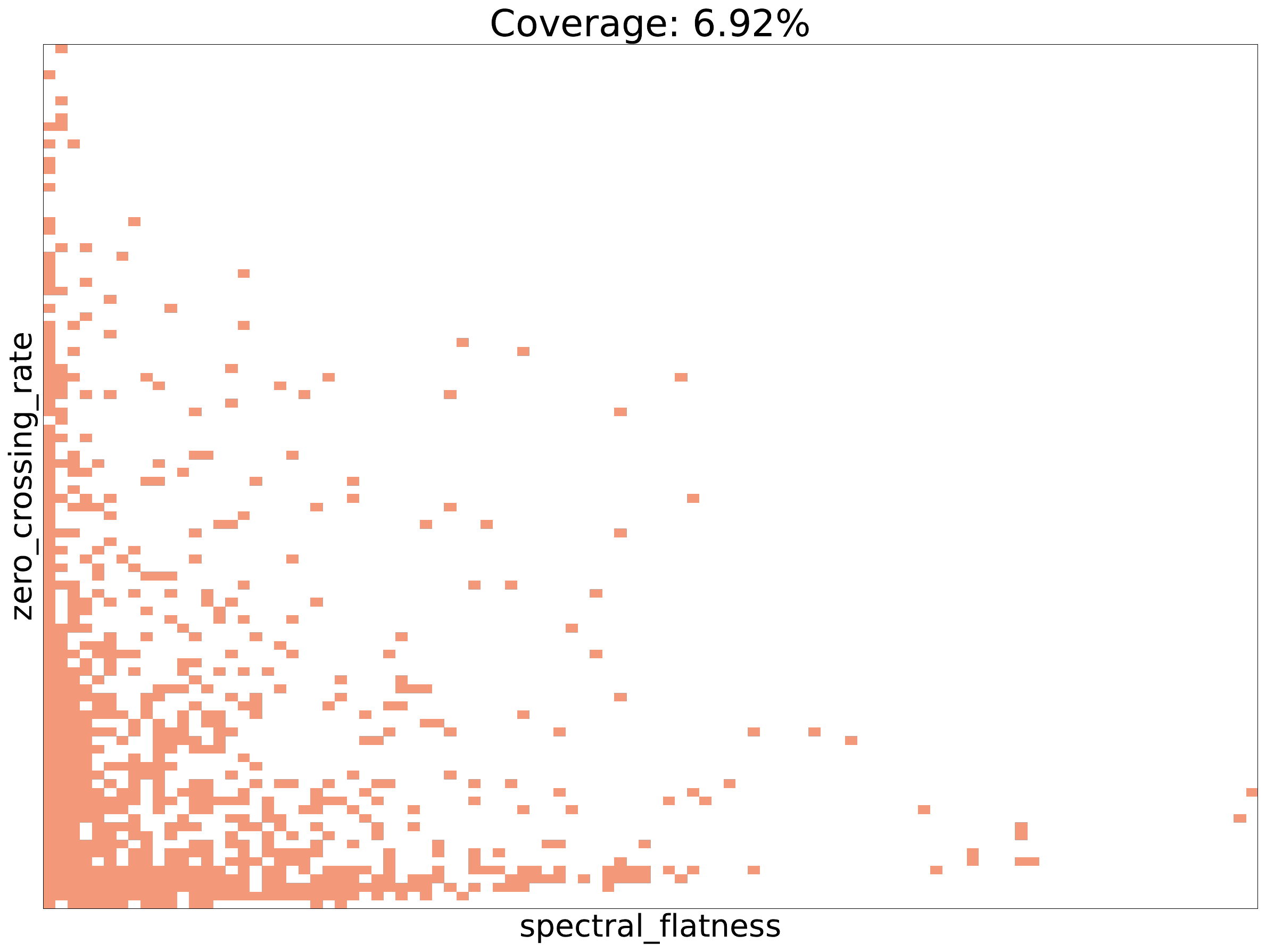}
    \end{tabular}
    \caption{Sounds from the NSynth dataset projected into behaviour spaces defined by combinations of instantaneous audio features, achieving varying percentages of coverage.}
    \label{fig:manual_bd_dataset_projections}
\end{figure*}

To support our choice of instantaneous audio features for manually defining behaviour spaces for \ac{QD} simulations, for comparison with our unsupervised approaches to such definitions, we projected sounds from the NSynth dataset to spaces delineated by several combinations of such features. Plots from such projections, to feature combinations that obtained the highest and lowest coverages, can be seen in Figure~\ref{fig:manual_bd_dataset_projections}. The difference between those two extremes shows the effect manual definition of the behaviour space can have, which we strive to free experimental designers from with the unsupervised approaches discussed in this paper.

\begin{acronym}[CPPN] %

\acro{EA}{Evolutionary Algorithm}
\acrodefplural{EA}[EAs]{Evolutionary Algorithms}

\acro{IEC}{Interactive Evolutionary Computation}

\acro{CPPN}{Compositional Pattern Producing Network}
\acrodefplural{CPPN}[CPPNs]{Compositional Pattern Producing Networks}

\acro{CSSN}{Compositional Sound Synthesis Network}
\acrodefplural{CSSN}[CSSNs]{Compositional Sound Synthesis Networks}

\acro{NEAT}{NeuroEvolution of Augmenting Topologies}
\acro{DSP}{Digital Signal Processing}

\acro{VAE}{Variational Auto Encoder}
\acrodefplural{VAE}[VAEs]{Variational Auto Encoders}

\acro{DNN}{Deep Neural Network}
\acrodefplural{DNN}[DNNs]{Deep Neural Network}

\acro{QD}{Quality Diversity}
\acro{MAP-Elites}{Multi-dimensional Archive of Phenotypic Elites }
\acro{NSLC}{Novelty Search with Local Competition}
\acro{YAMNet}{Yet Another Mobile Network}
\acro{HPC}{High Performance Computing}
\acro{gRPC}{gRPC Remote Procedure Calls}
\acro{WAV}{Waveform Audio File Format}

\acro{DR}{dimensionality reduction}

\acro{PCA}{Principal Component Analysis}
\acro{UMAP}{Uniform Manifold Approximation and Projection}
\acro{OEE}{Open-Ended Evolution}
\acro{OE}{Open-Endedness}
\acro{CC}{Computational Creativity}

\acro{BD}{Behavioural Descriptor}
\acrodefplural{BD}[BDs]{Behavioural Descriptors}

\acro{MFCC}{Mel-frequency cepstral coefficient}
\acrodefplural{MFCC}[MFCCs]{Mel-frequency cepstral coefficients}

\acro{DDSP}{Differentiable Digital Signal Processing}
\acro{FM}{frequency modulation}

\acro{HNSW}{Hierarchical Navigable Small World}

\acro{EDA}{Exploratory Data Analysis}

\acro{LLM}{large language model}
\acrodefplural{LLM}[LLMs]{large language model}

\end{acronym}

\end{document}